\documentclass[article]{achemso}

\usepackage{filecontents}
\usepackage{chemformula} % Formula subscripts using \ch{}
\usepackage[T1]{fontenc} % Use modern font encodings
\usepackage{natbib}
\usepackage{xr}
\usepackage[normalem]{ulem}
\usepackage{setspace}
\usepackage[mathletters]{ucs}
\usepackage[utf8x]{inputenc}

\author{Dominik~Stemer}
\affiliation[]{Fritz-Haber-Institut der Max-Planck-Gesellschaft, Berlin 14195, Germany.}
\email{dstemer@fhi-berlin.mpg.de}

\author{Stephan~Thürmer}
\affiliation[]{Department of Chemistry, Kyoto University, Kyoto 606-8502, Japan.}
\email{thuermer@kuchem.kyoto-u.ac.jp}

\author{Florian~Trinter}
\affiliation[]{Fritz-Haber-Institut der Max-Planck-Gesellschaft, Berlin 14195, Germany.}

\author{Uwe~Hergenhahn}
\affiliation[]{Fritz-Haber-Institut der Max-Planck-Gesellschaft, Berlin 14195, Germany.}

\author{Michele~Pugini}
\affiliation[]{Fritz-Haber-Institut der Max-Planck-Gesellschaft, Berlin 14195, Germany.}

\author{Bruno~Credidio}
\affiliation[]{Fritz-Haber-Institut der Max-Planck-Gesellschaft, Berlin 14195, Germany.}

\author{Sebastian~Malerz}
\affiliation[previous]{Fritz-Haber-Institut der Max-Planck-Gesellschaft, Berlin 14195, Germany.}
\altaffiliation{Department of Optics and Beamlines, Helmholtz-Zentrum Berlin für Materialien und Energie, Berlin 14109, Germany (present address).}

\author{Iain~Wilkinson}
\affiliation[]{Institute for Electronic Structure Dynamics, Helmholtz-Zentrum Berlin für Materialien und Energie, Berlin 14109, Germany.}

\author{Laurent~Nahon}
\affiliation[]{Synchrotron SOLEIL, St. Aubin 91190, France.}

\author{Gerard~Meijer}
\affiliation[]{Fritz-Haber-Institut der Max-Planck-Gesellschaft, Berlin 14195, Germany.}

\author{Ivan~Powis}
\affiliation[]{School of Chemistry, The University of Nottingham, Nottingham NG7 2RD, UK.}

\author{Bernd~Winter}
\affiliation[]{Fritz-Haber-Institut der Max-Planck-Gesellschaft, Berlin 14195, Germany.}
\email{winter@fhi-berlin.mpg.de}

\title[Photoelectron Circular Dichroism of Aqueous-Phase Alanine]{Photoelectron Circular Dichroism of Aqueous-Phase Alanine}

\abbreviations{PECD, LJ-PES}

\begin{document}

%%%%%%%%%%%%%%%%%%%%%%%%%%%%%%%%%%%%%%%%%%%%%%%%%%%%%%%%%%%%%%%%%%%%%
\begin{tocentry}

\includegraphics[width=8 cm]{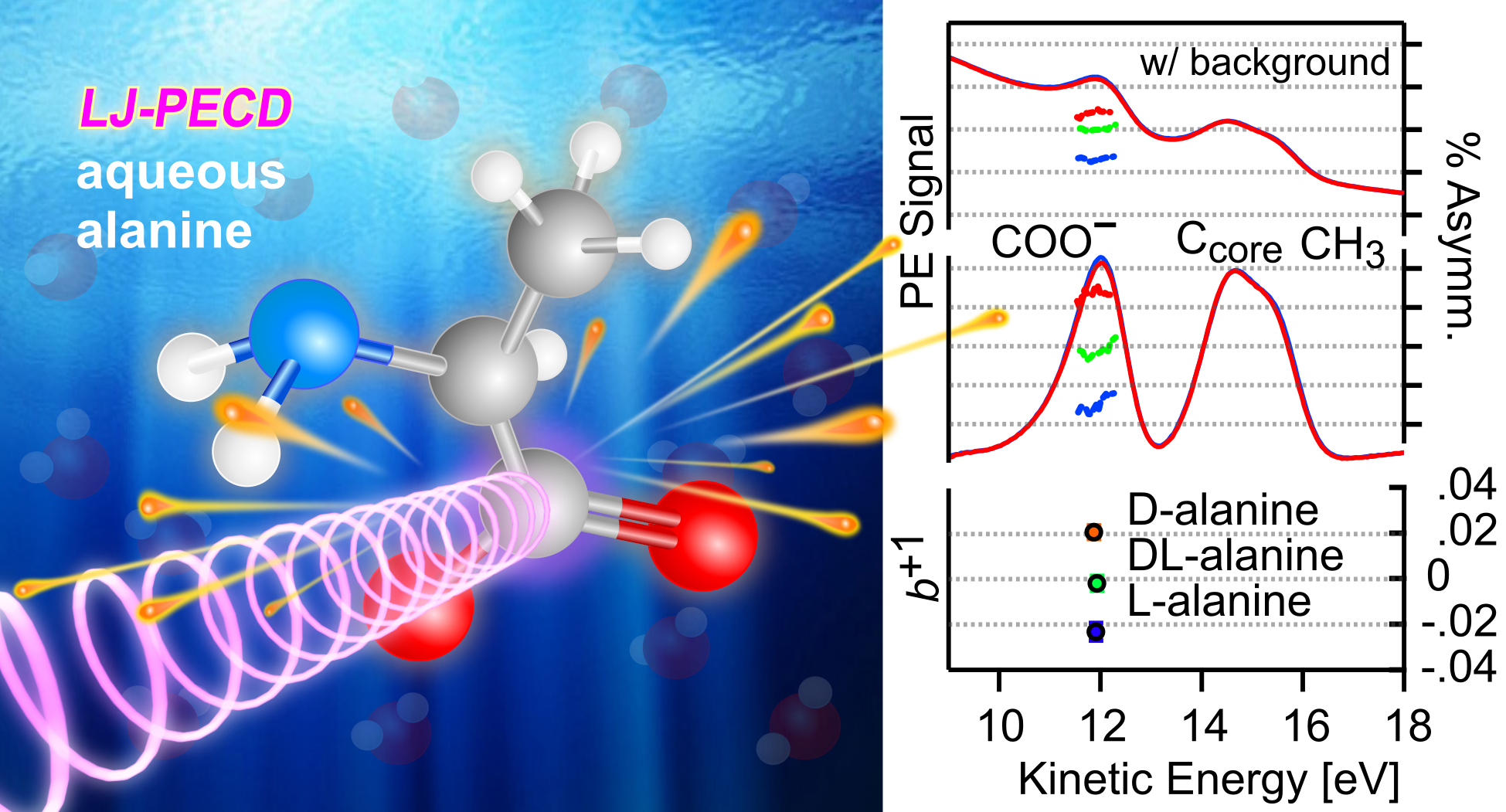}  

\end{tocentry}

%%%%%%%%%%%%%%%%%%%%%%%%%%%%%%%%%%%%%%%%%%%%%%%%%%%%%%%%%%%%%%%%%%%%%
\begin{abstract}
Amino acids and other small chiral molecules play key roles in biochemistry. However, in order to understand how these molecules behave in vivo, it is necessary to study them under aqueous-phase conditions. Photoelectron circular dichroism (PECD) has emerged as an extremely sensitive probe of chiral molecules, but its suitability for application to aqueous solutions had not yet been proven. Here, we report on our PECD measurements of aqueous-phase alanine, the simplest chiral amino acid. We demonstrate that the PECD response of alanine in water is different for each of alanine's carbon atoms, and is sensitive to molecular structure changes (protonation states) related to the solution pH. For C~1s photoionization of alanine's carboxylic acid group, we report PECD of comparable magnitude to that observed in valence-band photoelectron spectroscopy of gas-phase alanine. We identify key differences between PECD experiments from liquids and gases, discuss how PECD may provide information regarding solution-specific phenomena --- for example the nature and chirality of the solvation shell surrounding chiral molecules in water --- and highlight liquid-phase PECD as a powerful new tool for the study of aqueous-phase chiral molecules of biological relevance.
\end{abstract}

%%%%%%%%%%%%%%%%%%%%%%%%%%%%%%%%%%%%%%%%%%%%%%%%%%%%%%%%%%%%%%%%%%%%%
\section{Introduction}
Chirality describes a broad class of objects that share the property of being non-superposible with their mirror images. This general property of chirality may be found at all size scales and levels of complexity in nature, from human hands and sea shells down to small molecules. Many of the molecules most essential for life on Earth are chiral, including macromolecules such as proteins and DNA as well as smaller building blocks of life such as amino acids and sugars. All chiral biomolecules have been found to be overwhelmingly present in only a single enantiomeric form in living organisms, despite the mirror-image enantiomeric forms of a chiral molecule exhibiting identical physical and chemical properties in achiral environments (exempting small anticipated energetic differences between enantiomers due to parity-violating effects in the weak interactions) \cite{quack_high-resolution_2008}. The origins of this asymmetry in biology remain unresolved. In chiral environments, the chemical activity of the different enantiomers of a chiral molecule can differ dramatically. As such, experimental methods capable of distinguishing between enantiomers under biologically relevant conditions are of direct interest for chemical and life sciences. Absorption-based electronic circular dichroism (ECD) describes the asymmetry in absorption of left- or right-handed circularly polarized light (CPL) in the UV-visible range by different enantiomers of a chiral molecule. ECD is widely used as a means to determine the degree of enantiomeric excess for chiral molecules in solution \cite{berova_ECDreview_2007}. However, it is dependent on the interaction between the electric and magnetic dipoles of the chiral system with the CPL, and correspondingly the intrinsic effect magnitudes are small, generally on the order of 0.01\% \cite{nishino_eCD-pH2_2002}. 

In 2001, it was first experimentally demonstrated that one-photon photoionization of chiral molecules by circularly polarized synchrotron radiation may also be used to differentiate between enantiomers \cite{bowering_PECD_2001}. This photoelectron circular dichroism (PECD) --- predicted theoretically 25 years earlier by Ritchie \cite{ritchie_PECDtheory1, ritchie_PECDtheory2} --- manifests as a forward-backward asymmetry in the measured photoelectron flux along an experimental axis defined by the direction of light propagation. Arising purely due to electric-dipole interactions, PECD is much more pronounced than ECD, exhibiting asymmetries on the order of 1\% up to a few 10's\% for randomly oriented molecules \cite{nahon_PECDrev_2015}. Alignment of molecules has been demonstrated to further increase PECD effect sizes \cite{tia_alignedPECD_2017, fehre_fourfold_2021, nalin_molecular-frame_2023}. PECD appears to be a general feature in photoionization of chiral systems by CPL, having been observed in core-level and valence-band photoemission from terpenes \cite{hergenhahn_PECD_2004, nahon_valencecamphorPECD1_2006, powis_valencecamphorPECD2_2008, ulrich_giantPECD_2008} as well as from molecular dimers and larger clusters \cite{nahon_dimerPECD_2010, Powis_glycidol_clusters_2014}, amino acids \cite{tia_alaninePECD_2013, hartweg_condensation-PECD_2021, hadidi_conformer-dependentPECD_2021}, organometallic complexes \cite{darquie_Ru-AcAc-PECD_2021}, and nanoparticles \cite{hartweg_condensation-PECD_2021}. In addition to the traditional strengths of photoelectron spectroscopy, including chemical-state and site specificity, PECD is uniquely sensitive to small differences in molecular electronic structure and is capable of clearly distinguishing between different structural and conformational isomers of a chiral molecule \cite{turchini_conformerPECDrev_2017, rouquet_conformerselectivePECD_2024}. PECD is a powerful method capable of much more than the resolution of enantiomers and the determination of enantiomeric excess. These very appealing analytical capabilities have driven the growing field of laser-based multi-photon PECD since 2012.\cite{beaulieu_universalPECD_2016, lehmann_imagingPECD_2013, lux_circular_2012}

Water is critically important for biochemistry, playing an active role in determining the functionality of amino acids, proteins, and DNA through structure stabilization and mediation of intra- and intermolecular interactions \cite{levy_water-biochem_2006}. For chiral molecules in solution, interactions with solvent molecules, even achiral solvents such as water, may profoundly influence both the magnitude and the sign of measured chiroptical effects \cite{mennucci_modelingchirality_2011}. Additionally, it is possible that achiral solvent molecules may themselves adopt chiral arrangements within the first solvation shell around a chiral solute, mimicking to some extent the solute's chiral structure \cite{fidler_induced-eCD_1993}. Understanding the complex interactions between chiral solutes and achiral solvents is clearly a necessary prerequisite for developing deeper insight into the chemical activity of chiral molecules in solution, whether in biochemical or synthetic contexts. Core-level PECD seems uniquely suited for an investigation into the remarkably active role of solvent molecules in determining chiral solute properties. However, PECD of liquid-phase samples has only very recently become experimentally feasible \cite{malerz_EASI_2022, pohl_fenchone_PECD_2022}.

The development and growth of liquid-jet photoelectron spectroscopy (LJ-PES) over the past few decades has led to the solution of many of the technical obstacles that long prohibited the study of volatile liquids under the vacuum conditions required for PES \cite{winter_LJ-PES-review_2006, seidel_LJ-PES-Review_2016}. However, the application of PECD to liquids, and particularly to aqueous solutions, introduces a number of new experimental challenges that are not present for gas-phase studies. PECD, based upon the scattering and interference of departing partial electron waves, exhibits a strong kinetic-energy dependence, with the most pronounced effects manifesting for photoelectrons with kinetic energy of less than approximately 15~eV \cite{hergenhahn_PECD_2004}. For photoelectron spectroscopy of condensed-phase systems this poses a clear challenge due to the unavoidable convolution of primary photoelectron spectral features in this kinetic-energy range with the high-intensity low-energy electron background composed of electrons inelastically scattered in the sample bulk. For the cases of liquid water and of bulk-soluble species in aqueous solution, the opening of efficient quasi-elastic scattering channels within the same kinetic-energy range additionally leads to the broadening of photoelectron features and the homogenization of photoelectron angular distributions \cite{malerz_low-energy_2021, thurmer_waterPAD_2013}.

Despite these challenges, we have previously demonstrated the viability of liquid-phase PECD measurements for the chiral molecule fenchone as a neat liquid \cite{pohl_fenchone_PECD_2022}. We now report PECD in core-level photoionization of aqueous-phase alanine --- the simplest chiral amino acid --- at different solution pH values corresponding to the cationic, zwitterionic, and anionic form of the molecule. We summarize the results of our measurements, highlight the key differences in measuring PECD in the aqueous and gas phases, and discuss the potential influence of water solvation on PECD. We outline new opportunities for aqueous-phase PECD going forward, including the relatively straightforward application of PECD to study cationic or anionic chiral molecules; a topic that has only recently begun to be explored \cite{kruger_anionPECD_2021, kruger_anionPECD_2022, triptow_anionPECD2023}. More generally, this study constitutes the first core-level PECD investigation of chiral amino acids in any context. Liquid-phase PECD studies of the elementary constituents of life are highly relevant to in vivo biological conditions. We present a first step toward a bottom-up approach to understanding biomolecular complexity through the preparation of simple aqueous solutions, enabling the study of fragile thermolabile chiral species such as amino acids, peptides, sugars, and nucleic acids, whose vaporization into the gas phase with sufficient density to enable core-level PES is extremely challenging.

\subsection{Experimental Considerations}

\begin{figure} 
	\begin{center}
		\includegraphics[width=0.5\textwidth]{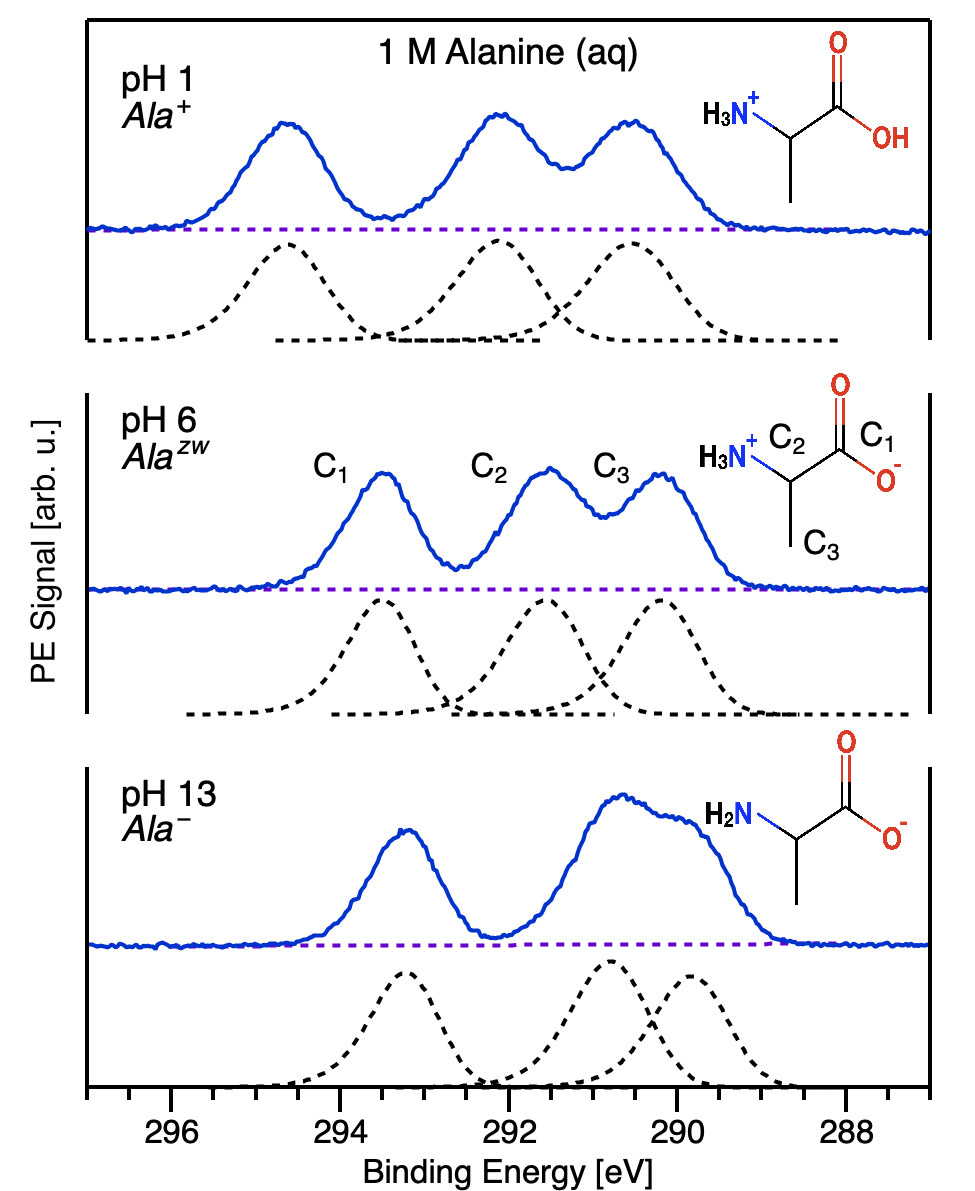}  
		\caption{Representative as-measured C~1s photoelectron spectra corresponding to aqueous-phase alanine's three chemically distinct carbons measured with $h\nu$~=~480.46~eV photons under acidic (top), neutral (middle), and basic (bottom) aqueous conditions. Measurements were conducted with 1~M aqueous solutions of L-alanine. The dashed black curves are fits to the data following the subtraction of background signals (dashed purple lines) using exponentially modified Gaussian profiles \cite{grushka_exmGauss_1972}. The asymmetry parameter was constrained to be between 0.2 and 0.3, and these values were kept constant in fits across different pH conditions. The ratios of peak widths were also constrained to differ by no more than 5\% across the three fits. Molecular structures of the dominant protonation state of alanine at the various pH values investigated are shown as insets.}
		\label{fig:XPS}
	\end{center}
\end{figure}

In this work, we focus on core-level C~1s measurements of aqueous-phase alanine. Alanine has three chemically distinct carbon atoms --- a carboxylic acid group, a chiral central carbon adjacent to an amine group, and a methyl group, which we denote C$_{1}$, C$_{2}$, and C$_{3}$, respectively --- resulting in a solution-phase C~1s PE spectrum with three distinct features that overlap to varying degrees depending on the solution pH / alanine protonation state (Fig.~\ref{fig:XPS}). 

To track chemical shifts upon protonation / deprotonation of alanine's functional groups, we measured C~1s PE spectra for aqueous solutions of alanine with solution pH values of 1, 6, and 13 using 480.46~eV photons. These pH values were chosen based on the $p\textit{K}_{a}$ values of alanine's functional groups (2.3 and 9.9 for alanine's carboxylic acid and amine groups in water, respectively) and ensured that, for a given solution pH, >95\% (>99\% for pH 6 and 13) of the alanine molecules in solution adopted a single charge state. For these measurements, we employed the energy-referencing scheme that we have developed over the past years.\cite{Thurmer_AccurateEnergies_2021, credidio_LJ-WF1_2022, pugini_LJ-WF2_2023} Briefly, this method enabled us to correct for the influence of electrokinetic charging and other stray fields by applying a suitable bias voltage to the liquid jet prior to measurement, thereby increasing the kinetic energy of all photoelectrons commensurately with the bias applied and revealing the secondary electron cutoff of the spectrum, which represents a true zero kinetic-energy reference for condensed-phase experiments. We note that it was not practical to employ these measures for the low-kinetic-energy PECD measurements discussed later, and thus the axes displayed in subsequent figures represent as-measured values, which may be subject to inaccuracies due to spectrometer offsets and the presence of streaming potentials and other stray fields \cite{kurahashi_streaming_2014,  winter_absKEreview_2023}.

Under strongly acidic conditions, alanine is predominantly in a cationic state ($Ala^{+}$), with the amine functional group protonated (Fig.~\ref{fig:XPS}, top). In this state, despite some overlap between C$_{2}$ and C$_{3}$, the three carbon PE features are easily resolvable owing to pronounced chemical shifts arising due to the different chemical environments of the carbon sites. The highest binding energy (BE) feature corresponds to photoelectrons originating from the relatively electronegative COOH group (C$_{1}$). The central PE feature corresponds to the chiral center adjacent to the amine group (C$_{2}$), and the final feature is attributable to the methyl group (C$_{3}$). These assignments are consistent with previous solid-state and gas-phase work \cite{clark_ala-solid_1976, powis_ala-gas_2003}. As can be seen by comparing the pH~1 and pH~6 data in Fig.~\ref{fig:XPS}, deprotonation of the carboxylic acid for zwitterionic Ala$^{zw}$ is accompanied by a noticeable chemical shift of the C$_{1}$ peak of approximately 1.12~eV toward lower BE, reflecting the increasing electron density at that site upon deprotonation. C$_{2}$ and C$_{3}$ also exhibit shifts to lower BE, albeit of smaller magnitudes, indicating their relatively lower sensitivity to the changing electron density on the neighboring functional group. This is likely a consequence of effective charge screening by nearby water molecules.\cite{credidio_proline_2024} At this pH, all three primary PE features are still readily resolvable. At pH~13, alanine is found in the anionic form, $Ala^{-}$, with the carboxylic acid deprotonated and the amine group neutral. In this case, we observe a chemical shift of 0.77~eV toward lower BE for the C$_{2}$ peak, corresponding to an increase in electron density at that group compared to the pH~6 case. This chemical shift results in substantial overlap of the C$_{2}$ and C$_{3}$ peaks. We note that the C$_{1}$ peak remains clearly isolated from the C$_{2}$ and C$_{3}$ features for all charge states, thereby simplifying the determination of PECD for this carbon site. The binding energies for C$_{1}$, C$_{2}$, and C$_{3}$ for solution pH 1, 6, and 13 are summarized in Table~\ref{tbl:XPSsummary}

Given the relatively high kinetic energy ($\approx190~eV$) of the photoelectrons measured, these C~1s measurements were not particularly surface sensitive. Nevertheless, by comparing the relative peak area of C$_{1}$, C$_{2}$, and C$_{3}$, we can draw initial inferences about the orientation of surface-localized molecules. Normalizing to the area of C$_{1}$, we find that the C$_{1}$:C$_{2}$:C$_{3}$ ratios are 1:1.09:1.13, 1:1.12:1.09, and 1:1.23:1.07 for Ala$^{+}$, Ala$^{zw}$, and Ala$^{-}$, respectively. Assuming that the C~1s photoionization cross section differs only negligibly for C$_{1}$, C$_{2}$, and C$_{3}$, a completely random molecular orientation distribution would result in a peak-area ratio of 1:1:1. A deviation from this ratio suggests partial molecular orientation for molecules near the liquid-vapor interface, with lower peak areas indicating that the photoionized groups are located deeper within the solution bulk \cite{bjorneholm_superficial_2022}. Our results suggest that C$_{1}$ appears to reside further in the solution bulk than C$_{2}$ or C$_{3}$ at all solution pH values probed. The disparity in the C$_{1}$:C$_{2}$ ratio is smallest for Ala$^{+}$ and most significant for Ala$^{-}$, where the charged carboxylic group is likely to be fully solvated. However, due to its overall hydrophilicty, alanine does not exhibit strong surface propensity \cite{mocellin_surface_2017}, and oriented molecules near the liquid-vapor interface are likely to make up only a fraction of the total molecules probed. Nevertheless, these potential effects are important to keep in mind, particularly for PECD measurements, which inevitably involve low-kinetic-energy electrons and are thus particularly surface sensitive.

\begin{table}[h]
\small
  \caption{C~1s binding energies for each of alanine's carbon groups as a function of solution pH.}
  \label{tbl:XPSsummary}
  \begin{tabular*}{0.6\textwidth}{@{\extracolsep{\fill}}llll}
    \hline
    Solution pH & BE C1 [eV] & BE C2 [eV] & BE C3 [eV]\\
    \hline
    1 & 294.64 & 292.12 & 290.53 \\
    6 & 293.52 & 291.56 & 290.19 \\
    13 & 293.25 & 290.79 & 289.81 \\
    \hline
  \end{tabular*}
\end{table}

Similarly to our previous study of PECD from neat liquid fenchone \cite{pohl_fenchone_PECD_2022}, we determined PECD using synchrotron radiation and an in-vacuum liquid-jet photoemission spectrometer via sequential measurements of photoelectron flux upon photoionization of aqueous alanine solutions with left- and right-handed CPL (see Experimental section below for additional details). The implemented beamline settings corresponding to each type of polarization were confirmed previously through measurements of PECD from gaseous fenchone \cite{malerz_EASI_2022, pohl_fenchone_PECD_2022}. Electrons were detected in the backward direction, with our electron analyzer mounted at $\approx$50$^\circ$ with respect to the axis of light propagation, approximating a magic-angle geometry of 54.74$^\circ$ \cite{malerz_EASI_2022}. This angle was chosen due to technical considerations of compatibility with the open port at the P04 soft x-ray beamline, where these measurements were performed.

For single-photon photoionization of a randomly oriented sample \cite{powis_PECDchapter_2008}, the normalized photoelectron angular distribution (PAD) within the electric-dipole approximation is given by:

\begin{equation}
	I^{p}(\theta) = 1 + b^{p}_{1}P_{1}(cos\theta) + b^{p}_{2}P_{2}(cos\theta)
	\label{Equation 1}
\end{equation}

\noindent Here, $I^{p}(\theta)$ is the photoionization intensity at an angle $\theta$ measured with respect to the photon-propagation direction for a given light polarization ($p$ = ±1 for circularly polarized light and 0 for linearly polarized light, respectively). $P_{n}$ represents the Legendre polynomial of the \textit{n}th order. The two coefficients $b^{p}_{1}$ and $b^{p}_{2}$ completely encompass the target-specific contribution to the PAD under a fixed set of experimental conditions. For achiral molecules, $b^{p}_{1} = 0$ and thus the PAD is described by $b^{p}_{2}$, more commonly denoted $\beta$ for experiments involving linearly polarized light (i.e. $b^{0}_{2} = \beta$, while $b^{\pm1}_{2} = -\beta/2$) \cite{reid_photoelectron_2003}. For chiral molecules, $b^{\pm 1}_{1}$ may be non-zero and $b^{+1}_{1} = -b^{-1}_{1}$, changing sign upon inversion of the handedness of CPL (or of the enantiomer). An asymmetry factor, $G$, may be defined as the simple difference between left- and right-handed PE intensities divided by their average. Using Eq. \ref{Equation 1} and noting that $P_{1}(x) = x$, we then have:

\begin{equation}
	\label{Equation 2}
	G=\frac{I^{+1}(\theta) - I^{-1}(\theta)}{[I^{+1}(\theta) + I^{-1}(\theta)]/2} = \frac{2b^{+1}_{1}cos\theta}{1+b^{\pm1}_{2}P_{2}(cos\theta)}
\end{equation}

It can be seen that this provides a means to extract $b^{+1}_{1}$ although, in general, a knowledge of $b^{\pm1}_{2}$ is also necessary. However, $P_{2}(cos(54.74^\circ)) = 0$. Thus, at magic angle:

\begin{equation}
	\label{Equation 3}
	b^{+1}_{1} = \pm\frac{[I^{+1} - I^{-1}]}{[I^{+1} + I^{-1}]} \times
	\frac{1}{cos(54.74^\circ)}	,	
\end{equation}

\noindent the negative sign applying when, as here, measurement is made in the backwards hemisphere. We note that $P_{2}(cos(50^\circ)) \neq 0$, and we discuss the relevant consequences of our near magic-angle measurement geometry in greater detail below.

As $b^{+1}_{1}$ simply changes sign upon exchange of enantiomer, in principle only a single pair of measurements for a single enantiomer with left- and right-handed CPL is necessary for its determination at any given energy using Eq.~\ref{Equation 2}. The normalized intensities $I^{p}(\theta)$ are obtained from the experimental photoelectron intensities at a specified electron energy and detection angle $\theta$, but require correction accounting for possible variation in photon flux and other sources of medium term experimental drift. Despite our best efforts to ensure that the fluxes of left- and right-handed CPL were equal --- monitored via a photodiode --- small persistent differences in light intensity of 1--3\% remained and so data normalization was essential. In gas-phase PECD studies, $4\pi$ imaging spectrometers enable collection of the total ionization yield, and thus enable direct normalization.\cite{nahon_valencecamphorPECD1_2006} In the present experimental arrangement the total ionization yield cannot be monitored and an alternative intensity-correction procedure utilizing the baseline signal was adopted. It is assumed that the background of scattered electrons is achiral and essentially independent of light polarization state. We therefore scaled all left- and right-handed CPL data (hereafter denoted as $+$ and $-$ spectra) to ensure equal intensity of the spectra at the low- and high-KE edges, where only an achiral background is present.

\begin{figure}
	\begin{center}
		\includegraphics[width=0.5\textwidth]{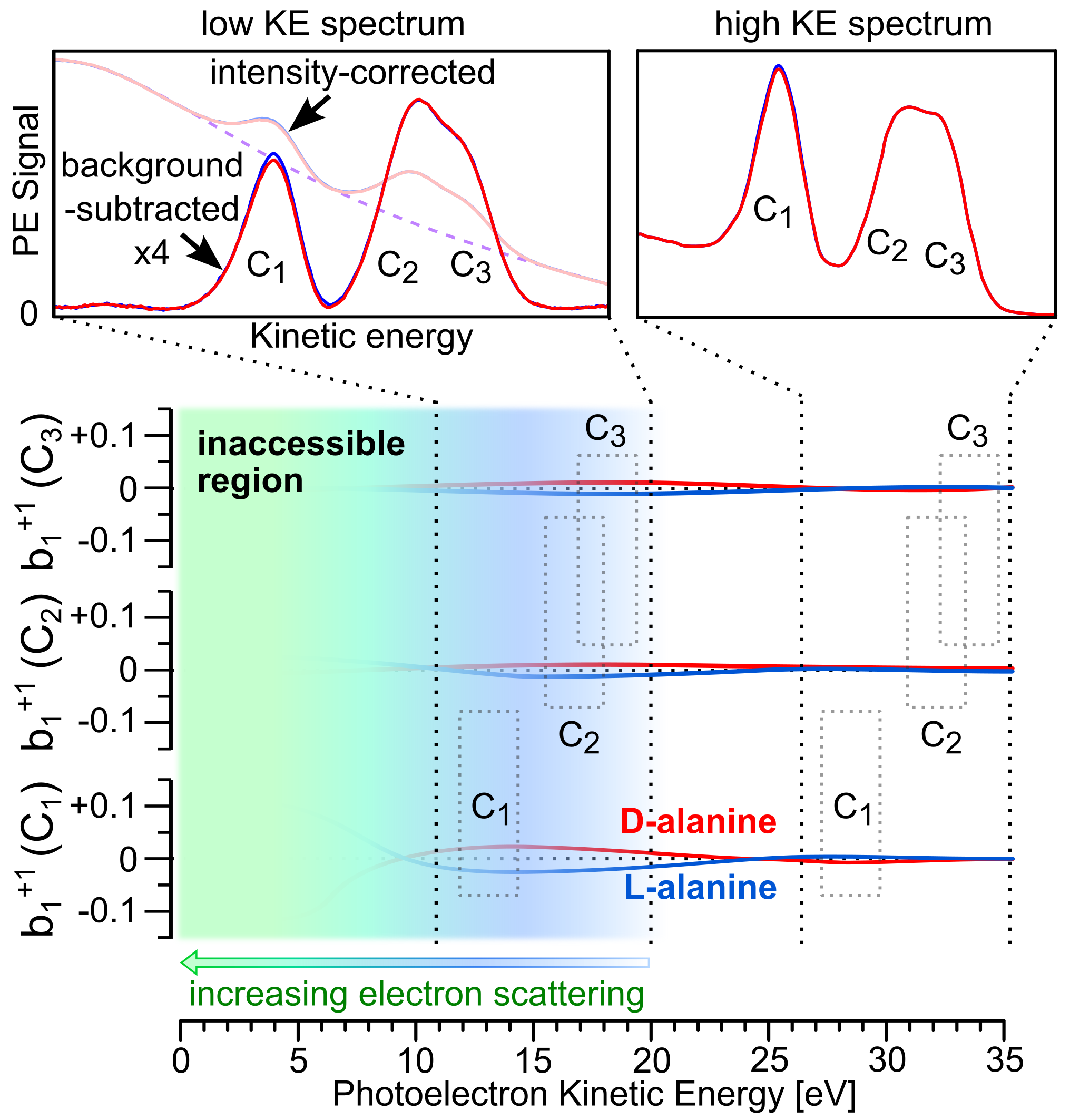} 
		\caption{Top: Illustrative aqueous-phase (pH 13) alanine C~1s photoelectron spectra corresponding to right- and left-handed circularly polarized light (red and blue lines, respectively). The high-energy spectra (right) exhibit low background signal, but vanishing photoelectron circular dichroism (PECD). The low-energy spectra sit atop a large background signal, but reveal significant PECD following subtraction of background signal (purple lines). Bottom: Hypothetical $b^{+1}_{1}$ parameters for each of alanine's carbon groups as a function of peak kinetic energy. The blue shaded zone represents the onset of significant kinetic-energy dependent elastic or quasi-elastic electron scattering. Within the green region, scattering of primary photoelectrons is sufficient to make resolution of the photoelectron features unfeasible. Values of $b^{+1}_{1}$ are currently inaccessible for this system within this kinetic-energy range. Dashed gray boxes represent approximate peak positions for the spectra shown above. The carbon atoms C$_{1}$, C$_{2}$, and C$_{3}$ are identified in Fig.~\ref{fig:XPS}. }
		\label{fig:Schema}
	\end{center}
\end{figure}

It is also vital that the baseline is accurately modeled such that the background beneath the peaks of interest can be reliably removed from the intensities $I$ used to determine $b^{+1}_1$. It can be seen from Eq.~\ref{Equation 2} that while any residual background present in the suitably scaled $I^+$ and $I^-$ will self-cancel in the numerator, this does not apply to the denominator, $[I^+ + I^-]$, whose magnitude would hence be overestimated, and lead to an erroneous underestimation of the derived $|b^{+1}_{1}|$. This challenge is exacerbated for experiments with dilute condensed-phase chiral molecules where the signal-to-background ratio is much reduced (see Fig.~\ref{fig:Schema}, top). Details regarding the background-subtraction method employed may be found in the supplementary information. 

\begin{figure}
	\begin{center}
		\includegraphics[width=0.5\textwidth]{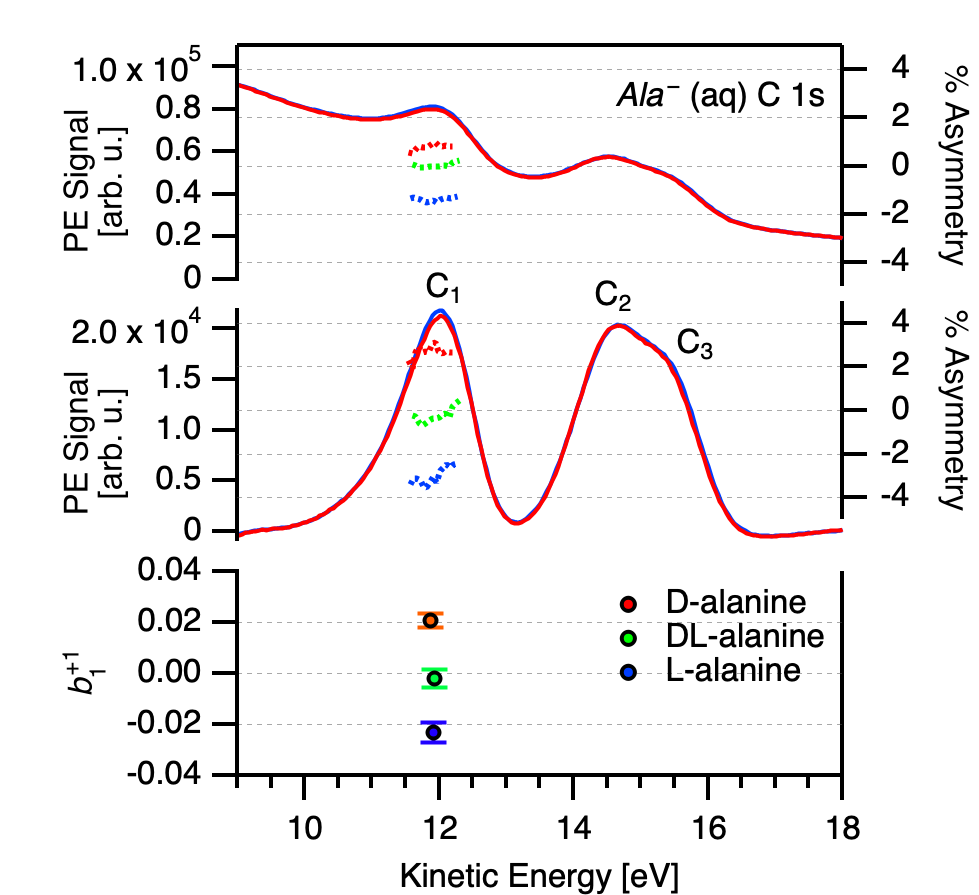} 
		\caption{Top: Representative intensity-scaled C~1s photoelectron spectra of 1~M L-alanine aqueous solution (pH~13) measured at $h\nu=305$~eV for left- and right-handed circularly polarized light (blue and red curves, respectively), along with the calculated percent difference between the spectra for C$_{1}$ group to background subtraction (right axis). Middle: Background-subtracted data, along with calculated percent difference between the spectra for C$_{1}$ (right axis). Bottom: Values of $b^{+1}_{1}$ obtained through the same process for D-, L-, and racemic DL-alanine (red, blue, and green points, respectively). The error bars of these points represent the standard deviation of the percent-difference data shown in the middle panel. The spectra shown here are averages of ten acquisitions, as typical for the data discussed in this study.}
		\label{fig:BGSub}
	\end{center}
\end{figure}

At the near magic-angle geometry of $\theta=50^\circ$ (in the backward direction) adopted for these experiments, the polynomial $P_2(\cos \theta) \neq 0$ and some residual influence of the $b^{\pm1}_{2}$ parameter is unavoidably included in the measurement (see Eq.~\ref{Equation 2}). However, there are currently no published values of $b^{\pm1}_{2}$ for core-level photoionization of alanine. To obtain an estimate of the significance of the remnant $b^{\pm1}_{2}$ contribution, we calculated the value of $b^{0}_{2}$ ($\beta$) for low-energy (KE = 0 to 25~eV) photoelectrons generated upon C 1s photoionization of gas-phase $Ala^{0}$ (details may be found in the supplementary information, see Fig.~\ref{SI-fig:S2}). We found that for photoionization of C$_1$, the absolute value of $\beta$ is less than 0.8 within this entire KE range, decreasing to less than 0.5 for photoelectrons with KE < 15~eV. Although aqueous-phase alanine will undoubtedly yield different values of $\beta$ due to differences in conformation and the charge state of its functional groups, this calculation nevertheless provides a rough estimate. As a comparison, the measured value of $\beta$ for O 1s photoionization of gas-phase water at 25~eV KE is approximately 1.6, whereas for liquid water at the same energy it is 0.6 \cite{thurmer_waterPAD_2013}. For both environments, $\beta$ trends strongly toward 0 for lower photoelectron kinetic energies. This decrease in $\beta$ as measured from liquid versus gaseous water was attributed primarily to highly efficient elastic or quasi-elastic scattering of photoelectrons by water molecules in the liquid. As water remains by far the likeliest scatterer of photoelectrons for the 1 M aqueous solutions of alanine studied in this work, it seems reasonable to expect a similar degree of PAD isotropization for $\beta$ as measured in C~1s photoionization of these solutions. Conservatively assuming a $\beta$ value of 0.5 as an upper limit for aqueous-phase alanine and using Eq.~\ref{Equation 2} while recalling that $b^{\pm 1}_2 = -\beta/2$ and simply averaging over the acceptance angle of the analyzer ($\approx \pm15^\circ$) \cite{malerz_EASI_2022}, the effect would be to scale our measured values of $b^{+1}_1$ by a factor of 0.97. Such a change does not alter the results of this study, and we do not consider it further in subsequent discussion.

For a given aqueous alanine solution (enantiomer, pH), we conducted C 1s LJ-PES measurements as a function of photon energy in the range of 302--310~eV, resulting in C$_{1}$ peaks in the KE range of 9–17 eV. For each photon energy, we measured 6–10 PE spectra before changing the polarization of the light, resulting in a measurement time of roughly 30--60 minutes per sample. We compared repeats using the same handedness of CPL for a given photon energy over time to ensure that our experimental conditions were reproducible during the measurement period. 

In practice, and particularly due to the pioneering nature of this study, it is advantageous to determine $b^{\pm1}_1$ for both enantiomers in order to detect any influence of instrumental and experimental asymmetry. Confirming  the anticipated enantiomeric mirroring of the $b^{+1}_{1}$ parameter assures the molecular origin of the observed asymmetry. Wherever possible, we also made the same observations for solutions of racemic DL-alanine, for which no asymmetry is expected. Due to the time-intensive nature of the experiments, which spanned multiple months of synchrotron measurements across four years, it was necessary to find a balance between the photoelectron KE range and enantiomer and protonation state of alanine investigated. This balance is reflected in the greater number of experimental data points measured for low-KE photoelectrons and for Ala$^{-}$ (see for reference Figs.~\ref{SI-fig:S3}, \ref{SI-fig:S7}, and \ref{SI-fig:S11}).

\section{Results}

\begin{figure}
	\begin{center}
		\includegraphics[width=\textwidth]{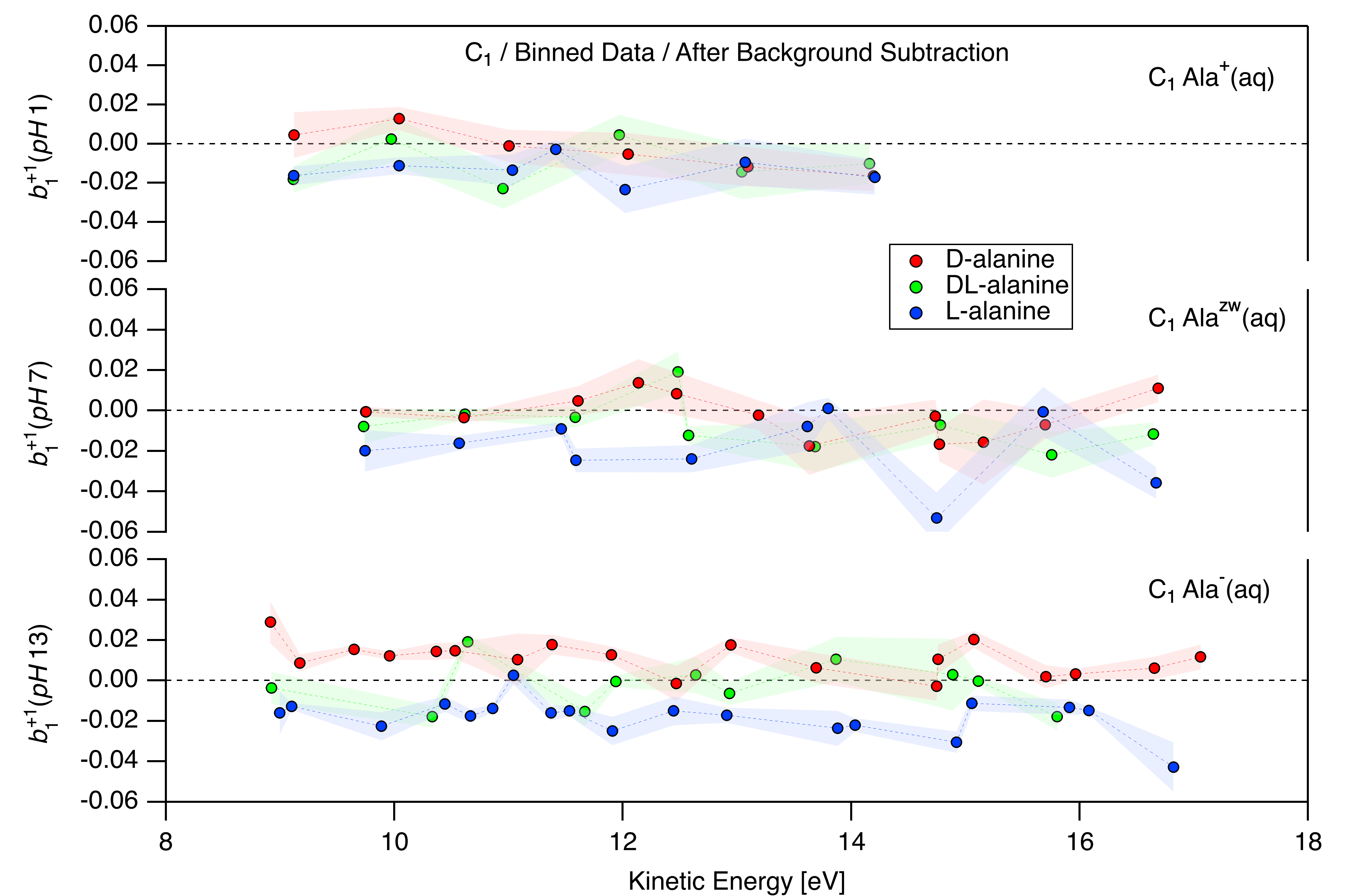}
		\caption{Values of the $b_{1}^{+1}$ photoionization parameter obtained for C~1s measurements of aqueous solutions of D-, L-, and DL-alanine (red, blue, and green points, respectively) at pH 1, 7, and 13 (top, middle, and bottom; corresponding to the cationic, zwitterionic, and anionic form of the molecule, respectively). All $b_{1}^{+1}$ values shown correspond to photoionization of the C$_{1}$ carboxylic group. The data displayed is the result of binning the results of individual measurement sets using a kinetic-energy window of 250 meV. The shaded error bars represent the combined error of the binned data points. A description of the error propagation may be found in the supplementary information.}
		\label{fig:Summary}
	\end{center}
\end{figure}

Following the process outlined in the supplementary information and using Eq.~\ref{Equation 2}, we determined a single value of $b^{+1}_{1}$ per carbon center for each pair of $+$ and $-$ measurements for a given enantiomer, solution pH, and photon energy (Fig.~\ref{fig:BGSub}). The data were then averaged within 250~meV KE windows for visualization, and to provide a realistic accuracy estimate for our measured photoelectron kinetic energies (see above for factors influencing the accuracy of "as-measured" kinetic energies). The complete datasets without averaging are presented in the supplementary information (Figs.~\ref{SI-fig:S4}, \ref{SI-fig:S8}, \ref{SI-fig:S12}). The analysis was carried out blind with respect to the enantiomer, such that we analyzed all the data without knowing whether any given pair of measurements corresponded to a solution of L-, D-, or DL-alanine. An overview of the $b^{+1}_{1}$ parameters obtained for C$_{1}$ at different pH values is presented in Fig.~\ref{fig:Summary}. Beginning with the pH~1 data, corresponding to $Ala^{+}$, weak enantiomer-dependent asymmetry in $b^{+1}_{1}$ may be seen for photoelectrons with KE 9~eV. At this energy, L-alanine exhibit a negative value of $b^{+1}_{1}$, while D-alanine yields a somewhat positive value. These values appear to rapidly trend toward zero at higher kinetic energies, and no significant PECD effect is found above 10~eV. For $Ala^{zw}$ at pH 6, the situation is different. Here, $b^{+1}_{1}$ is close to zero in the KE range of 9--11~eV, but appears to increase in magnitude briefly around 12--13~eV. From 14--16~eV, the data is noisy and does not indicate clear PECD. At 17~eV, $b^{+1}_{1}$ increases once more. The sign of $b^{+1}_{1}$ is the same for the corresponding enantiomer at pH~1. The clearest PECD can be seen for the case of $Ala^{-}$. For this protonation state, $b^{+1}_{1}$ as measured for C$_{1}$ is clearly non-zero for the entire KE range measured, even up to 17~eV. The sign of $b^{+1}_{1}$ remains constant throughout the entire range, and is consistent with that observed for $Ala^{zw}$ and $Ala^{+}$. Due to the very time-consuming nature of the experiments, it was not always possible to perform the same number of repeat measurements for D-, L-, and racemic alanine solutions. As such, it is critical to keep in mind that the points plotted in Fig.~\ref{fig:Summary} do not represent the same amount of measurements. This may be seen clearly for the unbinned data shown in Fig.~\ref{SI-fig:S4}.

We were unable to conclusively identify signatures of PECD for the C$_{2}$ and C$_{3}$ atoms, with the possible exception of $Ala^{zw}$ in the KE range of 15--19~eV (Figs.~\ref{SI-fig:S7} and \ref{SI-fig:S8}). However, due to the limited quantity of data in this KE range for $Ala^{zw}$ and the overlapping nature of these peaks, we do not consider the C$_{2}$ and C$_{3}$ data to convincingly demonstrate PECD. As mentioned above, we dedicated the most time to collecting data points for $Ala^{-}$, which appeared from the onset to yield the most clear PECD. As such, we are the most confident in our results for $Ala^{-}.$ It may seem surprising that C$_{1}$ exhibits clear PECD, while C$_{2}$ and C$_{3}$ do not. However, significant site-dependent PECD has previously been reported in C~1s studies of fenchone and glycidol,\cite{ulrich_giantPECD_2008, facciala_time-resolved-PECD_2023, Powis_glycidol_corePECD_2008} thereby highlighting the sensitivity of PECD to the specific atomic site probed. The datasets corresponding to Fig.\ref{fig:Summary} for C$_{2}$ and C$_{3}$ are available in the supplementary information (see Figs.~\ref{SI-fig:S7} and \ref{SI-fig:S11}).

\section{Discussion}

These results provide the only currently available report of PECD in aqueous-phase systems. They reveal a significant photoelectron circular dichroism (PECD) effect for aqueous-phase alanine in photoionization of the C$_1$ carbon group which falls within the range 0.5\% -- 40\% commonly encountered in gas-phase PECD studies. The sign of $b_1^{+1}$ is consistent across all three protonation states but there are differences in magnitude, with the most pronounced effects seen for $Ala^{-}$. Unlike many gas-phase PECD studies, we do not note a significant variation of the chiral asymmetry, $b_1^{+1}$, with photon energy.

Any PECD from the C$_2$ and C$_3$ photoemission essentially must fall below our presently achieved detection limits using a cylindrical liquid jet. Based upon previous experience with this arrangement we can estimate a possible reduction of any photoelectron anisotropy due to electron re-scattering within the bulk aqueous phase. In liquid water the angular anisotropy factor, $\beta$, of the O 1s PAD is reduced by 60\% compared to that of gaseous water at KE~=~25~eV \cite{thurmer_waterPAD_2013}. Similarly, in our previous core-level PECD study of liquid fenchone \cite{pohl_fenchone_PECD_2022}, we reported a factor of 5 decrease for $b_1^{+1}$ compared to that measured for gas-phase fenchone within the KE range of 9-13 eV \cite{ulrich_giantPECD_2008}. Hence, we tentatively estimate that our present results for $b_1^{+1}$ may be attenuated by a factor of 3--5. For the C$_1$ data shown in Fig.~4, this means that our measured values of $b_1^{+1} \approx 0.02$ may correspond to initial $b_1^{+1}$ values of approximately 0.06--0.1 prior to deleterious photoelectron scattering.

There is, \textit{a priori}, no particular surprise that the C$_1$ PECD should exceed that of the asymmetrically substituted C$_2$ site -- the so-called chiral center. The specific details, signs, and magnitudes of PECD are the result of extensive interference between the many continuum partial electron waves during photoionization, making any intuitive prediction or interpretation of signs and magnitudes extremely challenging. It is also well established that PECD asymmetry results from multiple scattering off the spatially extended chiral molecular framework \cite{hergenhahn_PECD_2004, ulrich_giantPECD_2008} and so is not solely contributed by a designated "chiral center". Indeed, strong PECD can be observed in molecules with no asymmetrically substituted centers but that display other chiral structural forms such as axial- or propeller-ligand symmetries. \cite{Rim_thesis, darquie_Ru-AcAc-PECD_2021}. Calculations for \textit{e.g.} fenchone \cite{facciala_time-resolved-PECD_2023} clearly show how the variation in $b_1^{+1}$ values over an extended kinetic-energy range mean that the relative signs and magnitudes observed from individual emitter sites will vary with observation energy.

PECD of gas-phase alanine has been reported in the valence region \cite{tia_alaninePECD_2013,tia_alaPECD2_2014} but
no C 1s measurements have been published so far. Were they available, simple comparisons would nevertheless be impeded by the different forms (ionic, zwitterionic) that alanine adopts in solution, and that we specifically targeted in this work. Then, even if these charge differences could be incorporated into current modeling, there is in general little value in comparing gas-phase measurements with those made in aqueous media without some understanding of further ways in which solvation may influence the observed PECD. Since our work aims to stimulate the development of theoretical modeling for liquid-phase PECD and, conversely, to show the extensive detail that may then be revealed with the benefit of comprehensive interpretive tools, we next set out further considerations of solvation effects. 

The scattering of emitted photoelectrons by bulk water has been alluded to above, but more intimate interactions between a molecule and its solvation shell require consideration. The PECD response is known to be very sensitive to molecular conformation, with great variation in magnitude and sign of the chiral asymmetry even with small changes such as the orientation of a terminal -OH rotamer \cite{Powis_glycidol_corePECD_2008, Garcia_glycidol_vpecd_2008}. This can be a hindrance or a benefit depending on one's perspective and purpose. Solvent--molecule interactions may change the molecular conformational landscape. Approximate treatments for solvation effects have been implemented, and are been widely applied, in many electronic-structure calculation packages. For example, the solvation shell of zwitterionic alanine and its consequent conformational rearrangement have been studied using such methods \cite{mullin_ala-water_2009}, but while these approaches may help identify preferred new conformations in a polar medium, they will lack explicit treatments of individual solvent molecules and their more specific interactions that can provide the level of insight likely to be required for accurate liquid-phase PECD calculations.

A study of microsolvated structures of protonated alanine \cite{Fischer_alanine_microsolvation} details how the addition of solvent molecules leads to the disruption of the intermolecular hydrogen bonds that stabilize and determine gas-phase conformers, replacing them with intramolecular molecule-solvent bonding until the first solvation shell completes with four attached water molecules. Such rearrangement, combined with stabilization of charge in the polar medium, can give rise to the phenomenon of conformational locking in solution, reducing the available conformational space as was inferred for the amino acid proline \cite{credidio_proline_2024}. On the other hand, conformational unlocking going from the gas phase to solution has been reported for asparagine \cite{Cabezas_asparagine_lock,Selvaraj_asparagine_unlock}, and molecular dynamics simulations of zwitterionic alanine in small water droplets have indicated that alanine does not adopt a single preferred conformer in aqueous solution \cite{degtyarenko_ala-water_2007}. Either an increase or a decrease in conformation flexibility could result in significant PECD changes, and would hence need careful identification. Even if a molecule were conformationally rigid, and so the chiral nuclear scattering potential relatively unaffected by solvation, there is evidence that PECD may be sensitive to the polarization of the electron density by H-bonding \cite{Powis_glycidol_clusters_2014}, and could thereby change when H-bonding was disrupted by solvation. 

Our results indicate that, even for molecules without significant surface activity, surface-orientation effects may not be trivial. Although it is currently not possible to correlate orientation of molecules at the liquid-vapor interfaces to observed PECD, given that orientation appears to increase the magnitude of PECD in gas-phase experiments \cite{tia_alignedPECD_2017, fehre_fourfold_2021, nalin_molecular-frame_2023}, explicit considerations of both surface propensity and molecular orientation seem to be necessary prerequisites for attempting to model solution-phase PECD.

The central importance of multiple scattering off the chiral molecular potential is particularly evident in core-level PECD when photemission occurs from an essentially spherical -- hence achiral -- 1s orbital. It is also commonly evident that such chiral final-state scattering may be a long range effect \cite{Harding_carvone_CMSXa_2006}. A number of investigations have examined PECD from weakly bound homochiral gas-phase clusters, of species as large as camphor dimers \cite{nahon_dimerPECD_2010}, that show changes in chiral angular distribution compared to the monomer. In a study of H-bonded \textit{n}-mers of glycidol, a total sign inversion of the strong monomer PECD was observed in the clustered species, with less pronounced differences between the individual $n=2, 3, 4$ clusters \cite{Powis_glycidol_clusters_2014}. More recently, a study of an achiral--chiral complex of phenol-methyloxirane \cite{rouquet_inducedPECD_2023} has shown evidence, namely photoelectron chiral asymmetry from ionization of the phenol chromophore, of induced chirality through such loose bonding. Such chirality transfer between solute and achiral solvent is well documented in, for example, vibrational circular dichroism (VCD)\cite{losada_vcd-h2o_2007, debie_induced-VCD_2008, yang_vcd-h2o-2_2009, jahnigen_non-covalent-induced_2021}. Hence, it seems inevitable that the detailed structure and interactions of at least the first solvation shell \cite{konstantinovsky_chiralSFG_2022, wang_MC_inducedCD_2008} would need to be incorporated into a realistic model of aqueous-phase PECD. In general, it appears likely that any effective model of PECD for solvated species must begin with a highly accurate representation of molecular conformation, hydrogen-bonding configuration, and molecular surface propensity and orientation.

One restriction on the currently achieved experimental sensitivity in this investigation stems from the lengthy data acquisition times per measurement and the consequent limitations on the statistical data quality. This is largely a consequence of the detection arrangement, whereby the electron analysis is performed by a hemispherical analyzer that can only measure at a small fixed solid angle and has to be scanned over the electron energy range of interest. Gas-phase PECD measurements were greatly advanced by the early adoption \cite{nahon_valencecamphorPECD1_2006} of velocity map imaging (VMI), which permits energy-resolved photoelectron angular distributions to be detected in a single measurement\cite{eppinkparker_vmi_1997}. The development of liquid-jet-compatible velocity map imaging is now an ongoing project, which we anticipate will dramatically decrease the amount of time needed for liquid-jet PECD measurements in the future, and thus significantly improve the general suitability of the technique for probing solution-phase molecular chirality.

\section{Conclusions}

We have measured core-level photoelectron circular dichroism (PECD) for aqueous-phase alanine, the smallest chiral amino acid, and examined each of its three charge states (cationic, zwitterionic, and anionic) by the variation of solution pH. A significant PECD was detected for photoionization of alanine's carboxylic acid functional group. This exhibited a clear magnitude dependence across pH conditions, but did not exhibit a strong kinetic-energy dependence. On the other hand, neither the photoionization of alanine's asymmetrically substituted  central carbon, nor its methyl carbon, displayed significant PECD, indicating that the effects for those groups are weaker and below our currently modest experimental sensitivity.

For photoionization of the carboxylic C 1s, PECD is the most pronounced under basic solution conditions, where alanine adopts the anionic form, but is very weak for both the zwitterionic and cationic forms. The differences observed across protonation states may be due to a number of factors, including changes in molecular charge state, conformation, surface propensity, molecular orientation, functional-group dependent solute-solvent interactions, and first solvation-shell structure.

Theoretical approaches to PECD are currently limited to small gas-phase systems and we identify some of the factors that may be required to develop realistically complete theoretical models for PECD effects in solution. Conversely, this discussion allows one to identify the scope of additional insight into the solvated state that could be exposed by the interpretation of PECD measurements. Experimentally, feasible technique developments of VMI to provide energy- and angle-multiplexed detection, coupled with the introduction of ultrathin liquid-jet technologies\cite{Koralek_ultrathin-FJ_2018} makes clear the potential for significantly improved data collection rates and enhanced detection sensitivity.

Our demonstration of aqueous-phase PECD marks a significant advance in the field of liquid-jet photoelectron spectroscopy. Given its chemical-state and site specificity and enantiomeric sensitivity, liquid-phase PECD is likely to become a remarkably useful analytical method for studying chiral molecules under biologically relevant aqueous conditions.

\section{Experimental}

We carried out all aqueous-phase PECD experiments using our custom LJ-PES apparatus --- Electronic Structure from Aqueous Solutions and Interfaces (EASI) --- developed specifically for this purpose \cite{malerz_EASI_2022}. We introduced alanine aqueous solutions into the experimental chamber using a custom quartz capillary, with an inner diameter on the order of 20--30~$\mu$m. Solutions were driven by a high-performance liquid chromatography pump (Shimadzu LC--20AD) at a flow rate of 0.8--1.5~mL/minute, resulting in solution velocities of 20--80~m/s. The solutions were prepared by diluting alanine in water, adjusting solution pH where necessary with concentrated HCl or solid NaOH, and finally diluting the solution to reach a final concentration of 1~M alanine. The pH~6 solutions represent as-prepared solutions, without the addition of either NaOH or HCl. The solutions were injected horizontally and frozen out after the interaction region via contact with a liquid-nitrogen-cooled cold trap. All PES experiments were performed at the P04 soft x-ray beamline of the PETRA~III storage ring at DESY in Hamburg, Germany. Critically, the P04 beamline is equipped with an APPLE-II undulator, enabling experiments with circularly polarized light within the experimentally relevant photon-energy range of 290-320~eV \cite{viefhaus_P04_2013}. Although not available during our early measurements, recent advancements at the P04 beamline now enable the estimation of the degree of circular polarization for a given set of undulator conditions. Retroactive estimates revealed an expected circular polarization of >99\% for all experimental conditions employed in this study. D-, L-, and DL-alanine (>98\% purity) were purchased from Sigma-Aldrich and Carl Roth and used as received. Enantiomeric purity was confirmed via intermittent testing using ECD (see Fig.~\ref{SI-fig:S1}). For additional experimental details regarding our instrumentation or LJ-PES more generally, we direct interested readers to our recent technical publication \cite{malerz_EASI_2022}. All data reduction was performed using the Igor Pro analysis software (Wavemetrics,version 9).

%%%%%%%%%%%%%%%%%%%%%%%%%%%%%%%%%%%%%%%%%%%%%%%%%%%%%%%%%%%%%%%%%%%%%

\begin{acknowledgement}
We acknowledge DESY (Hamburg, Germany), a member of the Helmholtz Association (HGF), for the provision of experimental facilities. Parts of this research were carried out at PETRA~III at the P04 soft x-ray beamline. Beamtime was allocated for proposals II-20180012, I-20200682, II-20210015, I-20211126, and I-20230378. We thank the P04 beamline staff, in particular Dr. Moritz Hoesch and Jörn Seltmann, for their long-standing support for this project. We also thank the European Molecular Biology Laboratory for access to the circular dichroism spectrophotometer. We thank Prof. Petr Slavíček, Prof. Philipp Demekhin, former group members Dr. Anne Stephansen, Dr. Claudia Kolbeck, Dr. Marvin Pohl, Dr. Karen Mudryk, and Dr. Tillmann Buttersack, as well as current group members Harmanjot Kaur, Qi Zhou, Dr. Nicolas Velasquez, Henrik Haak, and Sebastian Kray for valuable discussion and assistance with the transportation and preparation of our setup (EASI) for measurements at PETRA~III. We thank Dr. Lukáš Tomaník for support generally and for providing assistance with the representative C~1s measurements specifically.

D.S., U.H., M.P., B.C., and B.W. acknowledge funding from the European Research Council (ERC) under the European Union’s Horizon 2020 research and innovation program under Grant Agreement No. GAP 883759–AQUACHIRAL. F.T. acknowledges funding by the Deutsche Forschungsgemeinschaft (DFG, German Research Foundation) - Project 509471550, Emmy Noether Programme. F.T. and B.W. acknowledges support by the MaxWater initiative of the Max-Planck-Gesellschaft. B.W. acknowledges support by the Deutsche Forschungsgemeinschaft (Wi 1327/5-1). S.T. acknowledges support from ISHIZUE 2024 of Kyoto University.

\end{acknowledgement}

%%%%%%%%%%%%%%%%%%%%%%%%%%%%%%%%%%%%%%%%%%%%%%%%%%%%%%%%%%%%%%%%%%%%%
\begin{suppinfo}

The supplementary information contains electronic circular dichroism measurements of aqueous solutions of alanine, calculations of the photoionization parameter $b^{0}_{2}$ in the kinetic-energy range of 0--25~eV, a summary of all values of $b^{+1}_{1}$ measured for alanine's different carbon groups, including values calculated before background subtraction and kinetic-energy averaging was performed, and a description of the means determining the error of the $b^{+1}_{1}$ values, as shown in Figs.~\ref{fig:BGSub} and \ref{fig:Summary}.

The data of relevance to this study may be accessed at the following DOI: 10.5281/zenodo.13902253.
\end{suppinfo}

%%%%%%%%%%%%%%%%%%%%%%%%%%%%%%%%%%%%%%%%%%%%%%%%%%%%%%%%%%%%%%%%%%%%%
\bibliography{PECD_refs.bib}

\providecommand{\latin}[1]{#1}
\makeatletter
\providecommand{\doi}
  {\begingroup\let\do\@makeother\dospecials
  \catcode`\{=1 \catcode`\}=2 \doi@aux}
\providecommand{\doi@aux}[1]{\endgroup\texttt{#1}}
\makeatother
\providecommand*\mcitethebibliography{\thebibliography}
\csname @ifundefined\endcsname{endmcitethebibliography}
  {\let\endmcitethebibliography\endthebibliography}{}
\begin{mcitethebibliography}{72}
\providecommand*\natexlab[1]{#1}
\providecommand*\mciteSetBstSublistMode[1]{}
\providecommand*\mciteSetBstMaxWidthForm[2]{}
\providecommand*\mciteBstWouldAddEndPuncttrue
  {\def\EndOfBibitem{\unskip.}}
\providecommand*\mciteBstWouldAddEndPunctfalse
  {\let\EndOfBibitem\relax}
\providecommand*\mciteSetBstMidEndSepPunct[3]{}
\providecommand*\mciteSetBstSublistLabelBeginEnd[3]{}
\providecommand*\EndOfBibitem{}
\mciteSetBstSublistMode{f}
\mciteSetBstMaxWidthForm{subitem}{(\alph{mcitesubitemcount})}
\mciteSetBstSublistLabelBeginEnd
  {\mcitemaxwidthsubitemform\space}
  {\relax}
  {\relax}

\bibitem[Quack \latin{et~al.}(2008)Quack, Stohner, and
  Willeke]{quack_high-resolution_2008}
Quack,~M.; Stohner,~J.; Willeke,~M. High-resolution spectroscopic studies and
  theory of parity violation in chiral molecules. \emph{Annu. Rev. Phys. Chem.}
  \textbf{2008}, \emph{59}, 741--769\relax
\mciteBstWouldAddEndPuncttrue
\mciteSetBstMidEndSepPunct{\mcitedefaultmidpunct}
{\mcitedefaultendpunct}{\mcitedefaultseppunct}\relax
\EndOfBibitem
\bibitem[Berova \latin{et~al.}(2007)Berova, Bari, and
  Pescitelli]{berova_ECDreview_2007}
Berova,~N.; Bari,~L.~D.; Pescitelli,~G. Application of electronic circular
  dichroism in configurational and conformational analysis of organic
  compounds. \emph{Chemical Society Reviews} \textbf{2007}, \emph{36},
  914\relax
\mciteBstWouldAddEndPuncttrue
\mciteSetBstMidEndSepPunct{\mcitedefaultmidpunct}
{\mcitedefaultendpunct}{\mcitedefaultseppunct}\relax
\EndOfBibitem
\bibitem[Nishino \latin{et~al.}(2002)Nishino, Kosaka, Hembury, Matsushima, and
  Inoue]{nishino_eCD-pH2_2002}
Nishino,~H.; Kosaka,~A.; Hembury,~G.~A.; Matsushima,~K.; Inoue,~Y. The {pH}
  dependence of the anisotropy factors of essential amino acids. \emph{J. Chem.
  Soc., Perkin Trans. 2} \textbf{2002}, 582--590\relax
\mciteBstWouldAddEndPuncttrue
\mciteSetBstMidEndSepPunct{\mcitedefaultmidpunct}
{\mcitedefaultendpunct}{\mcitedefaultseppunct}\relax
\EndOfBibitem
\bibitem[Böwering \latin{et~al.}(2001)Böwering, Lischke, Schmidtke, Müller,
  Khalil, and Heinzmann]{bowering_PECD_2001}
Böwering,~N.; Lischke,~T.; Schmidtke,~B.; Müller,~N.; Khalil,~T.;
  Heinzmann,~U. Asymmetry in photoelectron emission from chiral molecules
  induced by circularly polarized light. \emph{Phys. Rev. Lett.} \textbf{2001},
  \emph{86}, 1187\relax
\mciteBstWouldAddEndPuncttrue
\mciteSetBstMidEndSepPunct{\mcitedefaultmidpunct}
{\mcitedefaultendpunct}{\mcitedefaultseppunct}\relax
\EndOfBibitem
\bibitem[Ritchie(1976)]{ritchie_PECDtheory1}
Ritchie,~B. Theory of the angular distribution of photoelectrons ejected from
  optically active molecules and molecular negative ions. \emph{Phys. Rev. A}
  \textbf{1976}, \emph{13}, 1411\relax
\mciteBstWouldAddEndPuncttrue
\mciteSetBstMidEndSepPunct{\mcitedefaultmidpunct}
{\mcitedefaultendpunct}{\mcitedefaultseppunct}\relax
\EndOfBibitem
\bibitem[Ritchie(1976)]{ritchie_PECDtheory2}
Ritchie,~B. Theory of angular distribution for ejection of photoelectrons from
  optically active molecules and molecular negative ions. {II}. \emph{Phys.
  Rev. A} \textbf{1976}, \emph{14}, 359\relax
\mciteBstWouldAddEndPuncttrue
\mciteSetBstMidEndSepPunct{\mcitedefaultmidpunct}
{\mcitedefaultendpunct}{\mcitedefaultseppunct}\relax
\EndOfBibitem
\bibitem[Nahon \latin{et~al.}(2015)Nahon, Garcia, and
  Powis]{nahon_PECDrev_2015}
Nahon,~L.; Garcia,~G.~A.; Powis,~I. Valence shell one-photon photoelectron
  circular dichroism in chiral systems. \emph{J Electron Spectrosc.}
  \textbf{2015}, \emph{204}, 322--334\relax
\mciteBstWouldAddEndPuncttrue
\mciteSetBstMidEndSepPunct{\mcitedefaultmidpunct}
{\mcitedefaultendpunct}{\mcitedefaultseppunct}\relax
\EndOfBibitem
\bibitem[Tia \latin{et~al.}(2017)Tia, Pitzer, Kastirke, Gatzke, Kim, Trinter,
  Rist, Hartung, Trabert, Siebert, Henrichs, Becht, Zeller, Gassert, Wiegandt,
  Wallauer, Kuhlins, Schober, Bauer, Wechselberger, Burzynski, Neff, Weller,
  Metz, Kircher, Waitz, Williams, Schmidt, Müller, Knie, Hans, Ben~Ltaief,
  Ehresmann, Berger, Fukuzawa, Ueda, Schmidt-Böcking, Dörner, Jahnke,
  Demekhin, and Schöffler]{tia_alignedPECD_2017}
Tia,~M. \latin{et~al.}  Observation of {Enhanced} {Chiral} {Asymmetries} in the
  {Inner}-{Shell} {Photoionization} of {Uniaxially} {Oriented} {Methyloxirane}
  {Enantiomers}. \emph{J. Phys. Chem. Lett.} \textbf{2017}, \emph{8},
  2780--2786\relax
\mciteBstWouldAddEndPuncttrue
\mciteSetBstMidEndSepPunct{\mcitedefaultmidpunct}
{\mcitedefaultendpunct}{\mcitedefaultseppunct}\relax
\EndOfBibitem
\bibitem[Fehre \latin{et~al.}(2021)Fehre, Novikovskiy, Grundmann, Kastirke,
  Eckart, Trinter, Rist, Hartung, Trabert, Janke, Nalin, Pitzer, Zeller,
  Wiegandt, Weller, Kircher, Hofmann, Schmidt, Knie, Hans, Ltaief, Ehresmann,
  Berger, Fukuzawa, Ueda, Schmidt-Böcking, Williams, Jahnke, Dörner,
  Schöffler, and Demekhin]{fehre_fourfold_2021}
Fehre,~K. \latin{et~al.}  Fourfold {Differential} {Photoelectron} {Circular}
  {Dichroism}. \emph{Phys. Rev. Lett.} \textbf{2021}, \emph{127}, 103201\relax
\mciteBstWouldAddEndPuncttrue
\mciteSetBstMidEndSepPunct{\mcitedefaultmidpunct}
{\mcitedefaultendpunct}{\mcitedefaultseppunct}\relax
\EndOfBibitem
\bibitem[Nalin \latin{et~al.}(2023)Nalin, Novikovskiy, Fehre, Anders, Trabert,
  Grundmann, Kircher, Khan, Tomar, Hofmann, Waitz, Vela-Perez, Kastirke,
  Siebert, Tsitsonis, Küstner-Wetekam, Marder, Viehmann, Trinter, Fukuzawa,
  Ueda, Williams, Knie, Dörner, Schöffler, Jahnke, and
  Demekhin]{nalin_molecular-frame_2023}
Nalin,~G. \latin{et~al.}  Molecular-frame differential photoelectron circular
  dichroism of {O} 1s-photoelectrons of trifluoromethyloxirane. \emph{Phys.
  Rev. Res.} \textbf{2023}, \emph{5}, 013021\relax
\mciteBstWouldAddEndPuncttrue
\mciteSetBstMidEndSepPunct{\mcitedefaultmidpunct}
{\mcitedefaultendpunct}{\mcitedefaultseppunct}\relax
\EndOfBibitem
\bibitem[Hergenhahn \latin{et~al.}(2004)Hergenhahn, Rennie, Kugeler, Marburger,
  Lischke, Powis, and Garcia]{hergenhahn_PECD_2004}
Hergenhahn,~U.; Rennie,~E.~E.; Kugeler,~O.; Marburger,~S.; Lischke,~T.;
  Powis,~I.; Garcia,~G. Photoelectron circular dichroism in core level
  ionization of randomly oriented pure enantiomers of the chiral molecule
  camphor. \emph{J. Chem. Phys.} \textbf{2004}, \emph{120}, 4553--4556\relax
\mciteBstWouldAddEndPuncttrue
\mciteSetBstMidEndSepPunct{\mcitedefaultmidpunct}
{\mcitedefaultendpunct}{\mcitedefaultseppunct}\relax
\EndOfBibitem
\bibitem[Nahon \latin{et~al.}(2006)Nahon, Garcia, Harding, Mikajlo, and
  Powis]{nahon_valencecamphorPECD1_2006}
Nahon,~L.; Garcia,~G.~A.; Harding,~C.~J.; Mikajlo,~E.; Powis,~I. Determination
  of chiral asymmetries in the valence photoionization of camphor enantiomers
  by photoelectron imaging using tunable circularly polarized light. \emph{J.
  Chem. Phys.} \textbf{2006}, \emph{125}, 114309\relax
\mciteBstWouldAddEndPuncttrue
\mciteSetBstMidEndSepPunct{\mcitedefaultmidpunct}
{\mcitedefaultendpunct}{\mcitedefaultseppunct}\relax
\EndOfBibitem
\bibitem[Powis \latin{et~al.}(2008)Powis, Harding, Garcia, and
  Nahon]{powis_valencecamphorPECD2_2008}
Powis,~I.; Harding,~C.~J.; Garcia,~G.~A.; Nahon,~L. A {Valence} {Photoelectron}
  {Imaging} {Investigation} of {Chiral} {Asymmetry} in the {Photoionization} of
  {Fenchone} and {Camphor}. \emph{ChemPhysChem} \textbf{2008}, \emph{9},
  475--483\relax
\mciteBstWouldAddEndPuncttrue
\mciteSetBstMidEndSepPunct{\mcitedefaultmidpunct}
{\mcitedefaultendpunct}{\mcitedefaultseppunct}\relax
\EndOfBibitem
\bibitem[Ulrich \latin{et~al.}(2008)Ulrich, Barth, Joshi, Hergenhahn, Mikajlo,
  Harding, and Powis]{ulrich_giantPECD_2008}
Ulrich,~V.; Barth,~S.; Joshi,~S.; Hergenhahn,~U.; Mikajlo,~E.; Harding,~C.~J.;
  Powis,~I. Giant {Chiral} {Asymmetry} in the {C} 1 \textit{s} {Core} {Level}
  {Photoemission} from {Randomly} {Oriented} {Fenchone} {Enantiomers}. \emph{J.
  Phys. Chem. A} \textbf{2008}, \emph{112}, 3544--3549\relax
\mciteBstWouldAddEndPuncttrue
\mciteSetBstMidEndSepPunct{\mcitedefaultmidpunct}
{\mcitedefaultendpunct}{\mcitedefaultseppunct}\relax
\EndOfBibitem
\bibitem[Nahon \latin{et~al.}(2010)Nahon, Garcia, Soldi-Lose, Daly, and
  Powis]{nahon_dimerPECD_2010}
Nahon,~L.; Garcia,~G.~A.; Soldi-Lose,~H.; Daly,~S.; Powis,~I. Effects of
  dimerization on the photoelectron angular distribution parameters from chiral
  camphor enantiomers obtained with circularly polarized vacuum-ultraviolet
  radiation. \emph{Phys. Rev. A} \textbf{2010}, \emph{82}, 032514\relax
\mciteBstWouldAddEndPuncttrue
\mciteSetBstMidEndSepPunct{\mcitedefaultmidpunct}
{\mcitedefaultendpunct}{\mcitedefaultseppunct}\relax
\EndOfBibitem
\bibitem[Powis \latin{et~al.}(2014)Powis, Daly, Tia, Cunha~de Miranda, Garcia,
  and Nahon]{Powis_glycidol_clusters_2014}
Powis,~I.; Daly,~S.; Tia,~M.; Cunha~de Miranda,~B.; Garcia,~G.~A.; Nahon,~L. A
  Photoionization Investigation of Small, Homochiral Clusters of Glycidol using
  Circularly Polarized Radiation and Velocity Map Electron-Ion Coincidence
  Imaging. \emph{Phys. Chem. Chem. Phys.} \textbf{2014}, \emph{16},
  467--476\relax
\mciteBstWouldAddEndPuncttrue
\mciteSetBstMidEndSepPunct{\mcitedefaultmidpunct}
{\mcitedefaultendpunct}{\mcitedefaultseppunct}\relax
\EndOfBibitem
\bibitem[Tia \latin{et~al.}(2013)Tia, Cunha De~Miranda, Daly, Gaie-Levrel,
  Garcia, Powis, and Nahon]{tia_alaninePECD_2013}
Tia,~M.; Cunha De~Miranda,~B.; Daly,~S.; Gaie-Levrel,~F.; Garcia,~G.~A.;
  Powis,~I.; Nahon,~L. Chiral asymmetry in the photoionization of gas-phase
  amino-acid alanine at lyman-alpha radiation wavelength. \emph{J. Phys. Chem.
  Lett.} \textbf{2013}, \emph{4}, 2698--2704\relax
\mciteBstWouldAddEndPuncttrue
\mciteSetBstMidEndSepPunct{\mcitedefaultmidpunct}
{\mcitedefaultendpunct}{\mcitedefaultseppunct}\relax
\EndOfBibitem
\bibitem[Hartweg \latin{et~al.}(2021)Hartweg, Garcia, Božanić, and
  Nahon]{hartweg_condensation-PECD_2021}
Hartweg,~S.; Garcia,~G.~A.; Božanić,~D.~K.; Nahon,~L. Condensation {Effects}
  on {Electron} {Chiral} {Asymmetries} in the {Photoionization} of {Serine}:
  {From} {Free} {Molecules} to {Nanoparticles}. \emph{J. Phys. Chem. Lett.}
  \textbf{2021}, \emph{12}, 2385--2393\relax
\mciteBstWouldAddEndPuncttrue
\mciteSetBstMidEndSepPunct{\mcitedefaultmidpunct}
{\mcitedefaultendpunct}{\mcitedefaultseppunct}\relax
\EndOfBibitem
\bibitem[Hadidi \latin{et~al.}(2021)Hadidi, Božanić, Ganjitabar, Garcia,
  Powis, and Nahon]{hadidi_conformer-dependentPECD_2021}
Hadidi,~R.; Božanić,~D.~K.; Ganjitabar,~H.; Garcia,~G.~A.; Powis,~I.;
  Nahon,~L. Conformer-dependent vacuum ultraviolet photodynamics and chiral
  asymmetries in pure enantiomers of gas phase proline. \emph{Commun. Chem.}
  \textbf{2021}, \emph{4}, 72\relax
\mciteBstWouldAddEndPuncttrue
\mciteSetBstMidEndSepPunct{\mcitedefaultmidpunct}
{\mcitedefaultendpunct}{\mcitedefaultseppunct}\relax
\EndOfBibitem
\bibitem[Darquié \latin{et~al.}(2021)Darquié, Saleh, Tokunaga, Srebro-Hooper,
  Ponzi, Autschbach, Decleva, Garcia, Crassous, and
  Nahon]{darquie_Ru-AcAc-PECD_2021}
Darquié,~B.; Saleh,~N.; Tokunaga,~S.~K.; Srebro-Hooper,~M.; Ponzi,~A.;
  Autschbach,~J.; Decleva,~P.; Garcia,~G.~A.; Crassous,~J.; Nahon,~L.
  Valence-shell photoelectron circular dichroism of
  ruthenium(iii)-tris-(acetylacetonato) gas-phase enantiomers. \emph{Phys.
  Chem. Chem. Phys.} \textbf{2021}, \emph{23}, 24140--24153\relax
\mciteBstWouldAddEndPuncttrue
\mciteSetBstMidEndSepPunct{\mcitedefaultmidpunct}
{\mcitedefaultendpunct}{\mcitedefaultseppunct}\relax
\EndOfBibitem
\bibitem[Turchini(2017)]{turchini_conformerPECDrev_2017}
Turchini,~S. Conformational effects in photoelectron circular dichroism.
  \emph{Journal of Physics: Condensed Matter} \textbf{2017}, \emph{29},
  503001\relax
\mciteBstWouldAddEndPuncttrue
\mciteSetBstMidEndSepPunct{\mcitedefaultmidpunct}
{\mcitedefaultendpunct}{\mcitedefaultseppunct}\relax
\EndOfBibitem
\bibitem[Rouquet \latin{et~al.}(2024)Rouquet, Dupont, Lepere, Garcia, Nahon,
  and Zehnacker]{rouquet_conformerselectivePECD_2024}
Rouquet,~E.; Dupont,~J.; Lepere,~V.; Garcia,~G.~A.; Nahon,~L.; Zehnacker,~A.
  Conformer-{Selective} {Photoelectron} {Circular} {Dichroism}. \emph{Angew.
  Chem. Int. Ed.} \textbf{2024}, \emph{63}, e202401423\relax
\mciteBstWouldAddEndPuncttrue
\mciteSetBstMidEndSepPunct{\mcitedefaultmidpunct}
{\mcitedefaultendpunct}{\mcitedefaultseppunct}\relax
\EndOfBibitem
\bibitem[Beaulieu \latin{et~al.}(2016)Beaulieu, Ferré, Géneaux, Canonge,
  Descamps, Fabre, Fedorov, Légaré, Petit, Ruchon, Blanchet, Mairesse, and
  Pons]{beaulieu_universalPECD_2016}
Beaulieu,~S.; Ferré,~A.; Géneaux,~R.; Canonge,~R.; Descamps,~D.; Fabre,~B.;
  Fedorov,~N.; Légaré,~F.; Petit,~S.; Ruchon,~T.; Blanchet,~V.; Mairesse,~Y.;
  Pons,~B. Universality of photoelectron circular dichroism in the
  photoionization of chiral molecules. \emph{New Journal of Physics}
  \textbf{2016}, \emph{18}, 102002\relax
\mciteBstWouldAddEndPuncttrue
\mciteSetBstMidEndSepPunct{\mcitedefaultmidpunct}
{\mcitedefaultendpunct}{\mcitedefaultseppunct}\relax
\EndOfBibitem
\bibitem[Lehmann \latin{et~al.}(2013)Lehmann, Ram, Powis, and
  Janssen]{lehmann_imagingPECD_2013}
Lehmann,~C.~S.; Ram,~N.~B.; Powis,~I.; Janssen,~M. H.~M. Imaging photoelectron
  circular dichroism of chiral molecules by femtosecond multiphoton coincidence
  detection. \emph{J. Chem. Phys.} \textbf{2013}, \emph{139}, 234307\relax
\mciteBstWouldAddEndPuncttrue
\mciteSetBstMidEndSepPunct{\mcitedefaultmidpunct}
{\mcitedefaultendpunct}{\mcitedefaultseppunct}\relax
\EndOfBibitem
\bibitem[Lux \latin{et~al.}(2012)Lux, Wollenhaupt, Bolze, Liang, Köhler,
  Sarpe, and Baumert]{lux_circular_2012}
Lux,~C.; Wollenhaupt,~M.; Bolze,~T.; Liang,~Q.; Köhler,~J.; Sarpe,~C.;
  Baumert,~T. Circular {Dichroism} in the {Photoelectron} {Angular}
  {Distributions} of {Camphor} and {Fenchone} from {Multiphoton} {Ionization}
  with {Femtosecond} {Laser} {Pulses}. \emph{Angew. Chem. Int. Ed}
  \textbf{2012}, \emph{51}, 5001--5005\relax
\mciteBstWouldAddEndPuncttrue
\mciteSetBstMidEndSepPunct{\mcitedefaultmidpunct}
{\mcitedefaultendpunct}{\mcitedefaultseppunct}\relax
\EndOfBibitem
\bibitem[Levy and Onuchic(2006)Levy, and Onuchic]{levy_water-biochem_2006}
Levy,~Y.; Onuchic,~J.~N. Water {Mediation} {In} {Protein} {Folding} and
  {Molecular} {Recognition}. \emph{Annu. Rev. Biophys.} \textbf{2006},
  \emph{35}, 389--415\relax
\mciteBstWouldAddEndPuncttrue
\mciteSetBstMidEndSepPunct{\mcitedefaultmidpunct}
{\mcitedefaultendpunct}{\mcitedefaultseppunct}\relax
\EndOfBibitem
\bibitem[Mennucci \latin{et~al.}(2011)Mennucci, Cappelli, Cammi, and
  Tomasi]{mennucci_modelingchirality_2011}
Mennucci,~B.; Cappelli,~C.; Cammi,~R.; Tomasi,~J. Modeling solvent effects on
  chiroptical properties. \emph{Chirality} \textbf{2011}, \emph{23},
  717--729\relax
\mciteBstWouldAddEndPuncttrue
\mciteSetBstMidEndSepPunct{\mcitedefaultmidpunct}
{\mcitedefaultendpunct}{\mcitedefaultseppunct}\relax
\EndOfBibitem
\bibitem[Fidler \latin{et~al.}(1993)Fidler, Rodger, and
  Rodger]{fidler_induced-eCD_1993}
Fidler,~J.; Rodger,~P.~M.; Rodger,~A. Circular dichroism as a probe of chiral
  solvent structure around chiral molecules. \emph{J. Chem. Soc., Perkin Trans.
  2} \textbf{1993}, 235--241\relax
\mciteBstWouldAddEndPuncttrue
\mciteSetBstMidEndSepPunct{\mcitedefaultmidpunct}
{\mcitedefaultendpunct}{\mcitedefaultseppunct}\relax
\EndOfBibitem
\bibitem[Malerz \latin{et~al.}(2022)Malerz, Haak, Trinter, Stephansen, Kolbeck,
  Pohl, Hergenhahn, Meijer, and Winter]{malerz_EASI_2022}
Malerz,~S.; Haak,~H.; Trinter,~F.; Stephansen,~A.~B.; Kolbeck,~C.; Pohl,~M.;
  Hergenhahn,~U.; Meijer,~G.; Winter,~B. A setup for studies of photoelectron
  circular dichroism from chiral molecules in aqueous solution. \emph{Rev. Sci.
  Instrum.} \textbf{2022}, \emph{93}, 015101\relax
\mciteBstWouldAddEndPuncttrue
\mciteSetBstMidEndSepPunct{\mcitedefaultmidpunct}
{\mcitedefaultendpunct}{\mcitedefaultseppunct}\relax
\EndOfBibitem
\bibitem[Pohl \latin{et~al.}(2022)Pohl, Malerz, Trinter, Lee, Kolbeck,
  Wilkinson, Thürmer, Neumark, Nahon, Powis, Meijer, Winter, and
  Hergenhahn]{pohl_fenchone_PECD_2022}
Pohl,~M.~N.; Malerz,~S.; Trinter,~F.; Lee,~C.; Kolbeck,~C.; Wilkinson,~I.;
  Thürmer,~S.; Neumark,~D.~M.; Nahon,~L.; Powis,~I.; Meijer,~G.; Winter,~B.;
  Hergenhahn,~U. Photoelectron circular dichroism in angle-resolved
  photoemission from liquid fenchone. \emph{Phys. Chem. Chem. Phys.}
  \textbf{2022}, \emph{24}, 8081--8092\relax
\mciteBstWouldAddEndPuncttrue
\mciteSetBstMidEndSepPunct{\mcitedefaultmidpunct}
{\mcitedefaultendpunct}{\mcitedefaultseppunct}\relax
\EndOfBibitem
\bibitem[Winter and Faubel(2006)Winter, and Faubel]{winter_LJ-PES-review_2006}
Winter,~B.; Faubel,~M. Photoemission from liquid aqueous solutions.
  \emph{Chemical Reviews} \textbf{2006}, \emph{106}, 1176--1211\relax
\mciteBstWouldAddEndPuncttrue
\mciteSetBstMidEndSepPunct{\mcitedefaultmidpunct}
{\mcitedefaultendpunct}{\mcitedefaultseppunct}\relax
\EndOfBibitem
\bibitem[Seidel \latin{et~al.}(2016)Seidel, Winter, and
  Bradforth]{seidel_LJ-PES-Review_2016}
Seidel,~R.; Winter,~B.; Bradforth,~S.~E. Valence Electronic Structure of
  Aqueous Solutions: Insights from Photoelectron Spectroscopy. \emph{Annu. Rev.
  Phys. Chem.} \textbf{2016}, \emph{67}, 283--305\relax
\mciteBstWouldAddEndPuncttrue
\mciteSetBstMidEndSepPunct{\mcitedefaultmidpunct}
{\mcitedefaultendpunct}{\mcitedefaultseppunct}\relax
\EndOfBibitem
\bibitem[Malerz \latin{et~al.}(2021)Malerz, Trinter, Hergenhahn, Ghrist, Ali,
  Nicolas, Saak, Richter, Hartweg, Nahon, Lee, Goy, Neumark, Meijer, Wilkinson,
  Winter, and Thürmer]{malerz_low-energy_2021}
Malerz,~S. \latin{et~al.}  Low-energy constraints on photoelectron spectra
  measured from liquid water and aqueous solutions. \emph{Phys. Chem. Chem.
  Phys.} \textbf{2021}, \emph{23}, 8246--8260\relax
\mciteBstWouldAddEndPuncttrue
\mciteSetBstMidEndSepPunct{\mcitedefaultmidpunct}
{\mcitedefaultendpunct}{\mcitedefaultseppunct}\relax
\EndOfBibitem
\bibitem[Th\"urmer \latin{et~al.}(2013)Th\"urmer, Seidel, Faubel, Eberhardt,
  Hemminger, Bradforth, and Winter]{thurmer_waterPAD_2013}
Th\"urmer,~S.; Seidel,~R.; Faubel,~M.; Eberhardt,~W.; Hemminger,~J.~C.;
  Bradforth,~S.~E.; Winter,~B. Photoelectron angular distributions from liquid
  water: {Effects} of electron scattering. \emph{Phys. Rev. Lett.}
  \textbf{2013}, \emph{111}, 173005\relax
\mciteBstWouldAddEndPuncttrue
\mciteSetBstMidEndSepPunct{\mcitedefaultmidpunct}
{\mcitedefaultendpunct}{\mcitedefaultseppunct}\relax
\EndOfBibitem
\bibitem[Krüger and Weitzel(2021)Krüger, and Weitzel]{kruger_anionPECD_2021}
Krüger,~P.; Weitzel,~K.-M. Photoelectron {Circular} {Dichroism} in the
  {Photodetachment} of {Amino} {Acid} {Anions}. \emph{Angew. Chem. Int. Ed.}
  \textbf{2021}, \emph{60}, 17861--17865\relax
\mciteBstWouldAddEndPuncttrue
\mciteSetBstMidEndSepPunct{\mcitedefaultmidpunct}
{\mcitedefaultendpunct}{\mcitedefaultseppunct}\relax
\EndOfBibitem
\bibitem[Krüger \latin{et~al.}(2022)Krüger, Both, Linne, Chirot, and
  Weitzel]{kruger_anionPECD_2022}
Krüger,~P.; Both,~J.~H.; Linne,~U.; Chirot,~F.; Weitzel,~K.-M. Photoelectron
  {Circular} {Dichroism} of {Electrosprayed} {Gramicidin} {Anions}. \emph{J.
  Phys. Chem. Lett.} \textbf{2022}, \emph{13}, 6110--6116\relax
\mciteBstWouldAddEndPuncttrue
\mciteSetBstMidEndSepPunct{\mcitedefaultmidpunct}
{\mcitedefaultendpunct}{\mcitedefaultseppunct}\relax
\EndOfBibitem
\bibitem[Triptow \latin{et~al.}(2023)Triptow, Fielicke, Meijer, and
  Green]{triptow_anionPECD2023}
Triptow,~J.; Fielicke,~A.; Meijer,~G.; Green,~M. Imaging {Photoelectron}
  {Circular} {Dichroism} in the {Detachment} of {Mass}-{Selected} {Chiral}
  {Anions}. \emph{Angew. Chem. Int. Ed.} \textbf{2023}, \emph{62},
  e202212020\relax
\mciteBstWouldAddEndPuncttrue
\mciteSetBstMidEndSepPunct{\mcitedefaultmidpunct}
{\mcitedefaultendpunct}{\mcitedefaultseppunct}\relax
\EndOfBibitem
\bibitem[Thurmer \latin{et~al.}(2021)Thurmer, Malerz, Trinter, Hergenhahn, Lee,
  Neumark, Meijer, Winter, and Wilkinson]{Thurmer_AccurateEnergies_2021}
Thurmer,~S.; Malerz,~S.; Trinter,~F.; Hergenhahn,~U.; Lee,~C.; Neumark,~D.~M.;
  Meijer,~G.; Winter,~B.; Wilkinson,~I. Accurate vertical ionization energy and
  work function determinations of liquid water and aqueous solutions.
  \emph{Chem. Sci.} \textbf{2021}, \emph{12}, 10558--10582\relax
\mciteBstWouldAddEndPuncttrue
\mciteSetBstMidEndSepPunct{\mcitedefaultmidpunct}
{\mcitedefaultendpunct}{\mcitedefaultseppunct}\relax
\EndOfBibitem
\bibitem[Credidio \latin{et~al.}(2022)Credidio, Pugini, Malerz, Trinter,
  Hergenhahn, Wilkinson, Thürmer, and Winter]{credidio_LJ-WF1_2022}
Credidio,~B.; Pugini,~M.; Malerz,~S.; Trinter,~F.; Hergenhahn,~U.;
  Wilkinson,~I.; Thürmer,~S.; Winter,~B. Quantitative electronic structure and
  work-function changes of liquid water induced by solute. \emph{Phys. Chem.
  Chem. Phys.} \textbf{2022}, \emph{24}, 1310--1325\relax
\mciteBstWouldAddEndPuncttrue
\mciteSetBstMidEndSepPunct{\mcitedefaultmidpunct}
{\mcitedefaultendpunct}{\mcitedefaultseppunct}\relax
\EndOfBibitem
\bibitem[Pugini \latin{et~al.}(2023)Pugini, Credidio, Walter, Malerz, Trinter,
  Stemer, Hergenhahn, Meijer, Wilkinson, Winter, and
  Thürmer]{pugini_LJ-WF2_2023}
Pugini,~M.; Credidio,~B.; Walter,~I.; Malerz,~S.; Trinter,~F.; Stemer,~D.;
  Hergenhahn,~U.; Meijer,~G.; Wilkinson,~I.; Winter,~B.; Thürmer,~S. How to
  measure work functions from aqueous solutions. \emph{Chem. Sci.}
  \textbf{2023}, \emph{14}, 9574--9588\relax
\mciteBstWouldAddEndPuncttrue
\mciteSetBstMidEndSepPunct{\mcitedefaultmidpunct}
{\mcitedefaultendpunct}{\mcitedefaultseppunct}\relax
\EndOfBibitem
\bibitem[Kurahashi \latin{et~al.}(2014)Kurahashi, Karashima, Tang, Horio,
  Abulimiti, Suzuki, Ogi, Oura, and Suzuki]{kurahashi_streaming_2014}
Kurahashi,~N.; Karashima,~S.; Tang,~Y.; Horio,~T.; Abulimiti,~B.;
  Suzuki,~Y.-I.; Ogi,~Y.; Oura,~M.; Suzuki,~T. Photoelectron spectroscopy of
  aqueous solutions: {Streaming} potentials of {NaX} ({X} = {Cl}, {Br}, and
  {I}) solutions and electron binding energies of liquid water and {X}
  $^{\textrm{−}}$. \emph{J. Chem. Phys.} \textbf{2014}, \emph{140},
  174506\relax
\mciteBstWouldAddEndPuncttrue
\mciteSetBstMidEndSepPunct{\mcitedefaultmidpunct}
{\mcitedefaultendpunct}{\mcitedefaultseppunct}\relax
\EndOfBibitem
\bibitem[Winter \latin{et~al.}(2023)Winter, Thürmer, and
  Wilkinson]{winter_absKEreview_2023}
Winter,~B.; Thürmer,~S.; Wilkinson,~I. Absolute {Electronic} {Energetics} and
  {Quantitative} {Work} {Functions} of {Liquids} from {Photoelectron}
  {Spectroscopy}. \emph{Acc. Chem. Res.} \textbf{2023}, \emph{56}, 77--85\relax
\mciteBstWouldAddEndPuncttrue
\mciteSetBstMidEndSepPunct{\mcitedefaultmidpunct}
{\mcitedefaultendpunct}{\mcitedefaultseppunct}\relax
\EndOfBibitem
\bibitem[Clark \latin{et~al.}(1976)Clark, Peeling, and
  Colling]{clark_ala-solid_1976}
Clark,~D.~T.; Peeling,~J.; Colling,~L. An experimental and theoretical
  investigation of the core level spectra of a series of amino acids,
  dipeptides and polypeptides. \emph{Biochim. Biophys. Acta} \textbf{1976},
  \emph{453}, 533--545\relax
\mciteBstWouldAddEndPuncttrue
\mciteSetBstMidEndSepPunct{\mcitedefaultmidpunct}
{\mcitedefaultendpunct}{\mcitedefaultseppunct}\relax
\EndOfBibitem
\bibitem[Powis \latin{et~al.}(2003)Powis, Rennie, Hergenhahn, Kugeler, and
  Bussy-Socrate]{powis_ala-gas_2003}
Powis,~I.; Rennie,~E.~E.; Hergenhahn,~U.; Kugeler,~O.; Bussy-Socrate,~R.
  Investigation of the gas-phase amino acid alanine by synchrotron radiation
  photoelectron spectroscopy. \emph{J. Phys. Chem. A} \textbf{2003},
  \emph{107}, 25--34\relax
\mciteBstWouldAddEndPuncttrue
\mciteSetBstMidEndSepPunct{\mcitedefaultmidpunct}
{\mcitedefaultendpunct}{\mcitedefaultseppunct}\relax
\EndOfBibitem
\bibitem[Credidio \latin{et~al.}(2024)Credidio, Thürmer, Stemer, Pugini,
  Trinter, Vokrouhlický, Slavíček, and Winter]{credidio_proline_2024}
Credidio,~B.; Thürmer,~S.; Stemer,~D.; Pugini,~M.; Trinter,~F.;
  Vokrouhlický,~J.; Slavíček,~P.; Winter,~B. From Gas to Solution: The
  Changing Neutral Structure of Proline upon Solvation. \emph{J. Phys. Chem. A}
  \textbf{2024}, \emph{128}, 10202--10212\relax
\mciteBstWouldAddEndPuncttrue
\mciteSetBstMidEndSepPunct{\mcitedefaultmidpunct}
{\mcitedefaultendpunct}{\mcitedefaultseppunct}\relax
\EndOfBibitem
\bibitem[Björneholm \latin{et~al.}(2022)Björneholm, Öhrwall, De~Brito,
  Ågren, and Carravetta]{bjorneholm_superficial_2022}
Björneholm,~O.; Öhrwall,~G.; De~Brito,~A.~N.; Ågren,~H.; Carravetta,~V.
  Superficial {Tale} of {Two} {Functional} {Groups}: {On} the {Surface}
  {Propensity} of {Aqueous} {Carboxylic} {Acids}, {Alkyl} {Amines}, and {Amino}
  {Acids}. \emph{Acc. Chem. Res.} \textbf{2022}, \emph{55}, 3285--3293\relax
\mciteBstWouldAddEndPuncttrue
\mciteSetBstMidEndSepPunct{\mcitedefaultmidpunct}
{\mcitedefaultendpunct}{\mcitedefaultseppunct}\relax
\EndOfBibitem
\bibitem[Mocellin \latin{et~al.}(2017)Mocellin, Gomes, Araújo, De~Brito, and
  Björneholm]{mocellin_surface_2017}
Mocellin,~A.; Gomes,~A. H. D.~A.; Araújo,~O.~C.; De~Brito,~A.~N.;
  Björneholm,~O. Surface {Propensity} of {Atmospherically} {Relevant} {Amino}
  {Acids} {Studied} by {XPS}. \emph{J. Phys. Chem. B} \textbf{2017},
  \emph{121}, 4220--4225\relax
\mciteBstWouldAddEndPuncttrue
\mciteSetBstMidEndSepPunct{\mcitedefaultmidpunct}
{\mcitedefaultendpunct}{\mcitedefaultseppunct}\relax
\EndOfBibitem
\bibitem[Grushka(1972)]{grushka_exmGauss_1972}
Grushka,~E. Characterization of exponentially modified {Gaussian} peaks in
  chromatography. \emph{Analytical Chemistry} \textbf{1972}, \emph{44},
  1733--1738\relax
\mciteBstWouldAddEndPuncttrue
\mciteSetBstMidEndSepPunct{\mcitedefaultmidpunct}
{\mcitedefaultendpunct}{\mcitedefaultseppunct}\relax
\EndOfBibitem
\bibitem[Powis(2008)]{powis_PECDchapter_2008}
Powis,~I. In \emph{Advances in {Chemical} {Physics}}; Rice,~S.~A., Ed.; John
  Wiley \& Sons, Inc.: Hoboken, NJ, USA, 2008; Vol. 138; pp 267--329\relax
\mciteBstWouldAddEndPuncttrue
\mciteSetBstMidEndSepPunct{\mcitedefaultmidpunct}
{\mcitedefaultendpunct}{\mcitedefaultseppunct}\relax
\EndOfBibitem
\bibitem[Reid(2003)]{reid_photoelectron_2003}
Reid,~K.~L. Photoelectron angular distributions. \emph{Annu. Rev. Phys. Chem.}
  \textbf{2003}, \emph{54}, 397--424\relax
\mciteBstWouldAddEndPuncttrue
\mciteSetBstMidEndSepPunct{\mcitedefaultmidpunct}
{\mcitedefaultendpunct}{\mcitedefaultseppunct}\relax
\EndOfBibitem
\bibitem[Facciala \latin{et~al.}(2023)Facciala, Devetta, Beauvarlet, Besley,
  Calegari, Callegari, Catone, Cinquanta, Ciriolo, Colaizzi, Coreno, Crippa,
  De~Ninno, Di~Fraia, Galli, Garcia, Mairesse, Negro, Plekan, Prasannan~Geetha,
  Prince, Pusala, Stagira, Turchini, Ueda, You, Zema, Blanchet, Nahon, Powis,
  and Vozzi]{facciala_time-resolved-PECD_2023}
Facciala,~D. \latin{et~al.}  Time-{Resolved} {Chiral} {X}-{Ray} {Photoelectron}
  {Spectroscopy} with {Transiently} {Enhanced} {Atomic} {Site} {Selectivity}:
  {A} {Free}-{Electron} {Laser} {Investigation} of {Electronically} {Excited}
  {Fenchone} {Enantiomers}. \emph{Phys. Rev. X} \textbf{2023}, \emph{13},
  011044\relax
\mciteBstWouldAddEndPuncttrue
\mciteSetBstMidEndSepPunct{\mcitedefaultmidpunct}
{\mcitedefaultendpunct}{\mcitedefaultseppunct}\relax
\EndOfBibitem
\bibitem[Powis \latin{et~al.}(2008)Powis, Harding, Barth, Joshi, Ulrich, and
  Hergenhahn]{Powis_glycidol_corePECD_2008}
Powis,~I.; Harding,~C.~J.; Barth,~S.; Joshi,~S.; Ulrich,~V.; Hergenhahn,~U.
  Chiral Asymmetry in the Angle-Resolved O and C 1s-1 Core Photoemissions of
  the R Enantiomer of Glycidol. \emph{Phys. Rev. A} \textbf{2008}, \emph{78},
  052501\relax
\mciteBstWouldAddEndPuncttrue
\mciteSetBstMidEndSepPunct{\mcitedefaultmidpunct}
{\mcitedefaultendpunct}{\mcitedefaultseppunct}\relax
\EndOfBibitem
\bibitem[Hadidi(2020)]{Rim_thesis}
Hadidi,~R. Dichroïsme Circulaire de Photoélectron (PECD) en couche de valence
  : systèmes à base d’acides aminés et dérivés de binaphtyles. Ph.D.\
  thesis, Université Paris-Saclay, 2020\relax
\mciteBstWouldAddEndPuncttrue
\mciteSetBstMidEndSepPunct{\mcitedefaultmidpunct}
{\mcitedefaultendpunct}{\mcitedefaultseppunct}\relax
\EndOfBibitem
\bibitem[Tia \latin{et~al.}(2014)Tia, Cunha De~Miranda, Daly, Gaie-Levrel,
  Garcia, Nahon, and Powis]{tia_alaPECD2_2014}
Tia,~M.; Cunha De~Miranda,~B.; Daly,~S.; Gaie-Levrel,~F.; Garcia,~G.~A.;
  Nahon,~L.; Powis,~I. {VUV} photodynamics and chiral asymmetry in the
  photoionization of gas phase alanine enantiomers. \emph{J. Phys. Chem. A}
  \textbf{2014}, \emph{118}, 2765--2779\relax
\mciteBstWouldAddEndPuncttrue
\mciteSetBstMidEndSepPunct{\mcitedefaultmidpunct}
{\mcitedefaultendpunct}{\mcitedefaultseppunct}\relax
\EndOfBibitem
\bibitem[Garcia \latin{et~al.}(2008)Garcia, Nahon, Harding, and
  Powis]{Garcia_glycidol_vpecd_2008}
Garcia,~G.~A.; Nahon,~L.; Harding,~C.~J.; Powis,~I. Chiral Signatures in Angle
  Resolved Valence Photoelectron Spectroscopy of Pure Glycidol Enantiomers.
  \emph{Phys. Chem. Chem. Phys.} \textbf{2008}, \emph{10}, 1628--1639\relax
\mciteBstWouldAddEndPuncttrue
\mciteSetBstMidEndSepPunct{\mcitedefaultmidpunct}
{\mcitedefaultendpunct}{\mcitedefaultseppunct}\relax
\EndOfBibitem
\bibitem[Mullin and Gordon(2009)Mullin, and Gordon]{mullin_ala-water_2009}
Mullin,~J.~M.; Gordon,~M.~S. Alanine: {Then} {There} {Was} {Water}. \emph{J.
  Phys. Chem. B} \textbf{2009}, \emph{113}, 8657--8669\relax
\mciteBstWouldAddEndPuncttrue
\mciteSetBstMidEndSepPunct{\mcitedefaultmidpunct}
{\mcitedefaultendpunct}{\mcitedefaultseppunct}\relax
\EndOfBibitem
\bibitem[Fischer \latin{et~al.}(2019)Fischer, Sherman, Voss, Zhou, and
  Garand]{Fischer_alanine_microsolvation}
Fischer,~K.~C.; Sherman,~S.~L.; Voss,~J.~M.; Zhou,~J.; Garand,~E.
  Microsolvation Structures of Protonated Glycine and l-Alanine. \emph{J. Phys.
  Chem. A} \textbf{2019}, \emph{123}, 3355--3366\relax
\mciteBstWouldAddEndPuncttrue
\mciteSetBstMidEndSepPunct{\mcitedefaultmidpunct}
{\mcitedefaultendpunct}{\mcitedefaultseppunct}\relax
\EndOfBibitem
\bibitem[Cabezas \latin{et~al.}(2012)Cabezas, Varela, Peña, Mata, López, and
  Alonso]{Cabezas_asparagine_lock}
Cabezas,~C.; Varela,~M.; Peña,~I.; Mata,~S.; López,~J.~C.; Alonso,~J.~L. The
  conformational locking of asparagine. \emph{Chem. Commun.} \textbf{2012},
  \emph{48}, 5934--5936\relax
\mciteBstWouldAddEndPuncttrue
\mciteSetBstMidEndSepPunct{\mcitedefaultmidpunct}
{\mcitedefaultendpunct}{\mcitedefaultseppunct}\relax
\EndOfBibitem
\bibitem[Selvaraj \latin{et~al.}(2012)Selvaraj, Murugan, and
  Ågren]{Selvaraj_asparagine_unlock}
Selvaraj,~A. R.~K.; Murugan,~N.~A.; Ågren,~H. Solvent Polarity-Induced
  Conformational Unlocking of Asparagine. \emph{J. Phys. Chem. A}
  \textbf{2012}, \emph{116}, 11702--11708\relax
\mciteBstWouldAddEndPuncttrue
\mciteSetBstMidEndSepPunct{\mcitedefaultmidpunct}
{\mcitedefaultendpunct}{\mcitedefaultseppunct}\relax
\EndOfBibitem
\bibitem[Degtyarenko \latin{et~al.}(2007)Degtyarenko, Jalkanen, Gurtovenko, and
  Nieminen]{degtyarenko_ala-water_2007}
Degtyarenko,~I.~M.; Jalkanen,~K.~J.; Gurtovenko,~A.~A.; Nieminen,~R.~M. Alanine
  in a {Droplet} of {Water}: {A} {Density}-{Functional} {Molecular} {Dynamics}
  {Study}. \emph{J. Phys. Chem. B} \textbf{2007}, \emph{111}, 4227--4234\relax
\mciteBstWouldAddEndPuncttrue
\mciteSetBstMidEndSepPunct{\mcitedefaultmidpunct}
{\mcitedefaultendpunct}{\mcitedefaultseppunct}\relax
\EndOfBibitem
\bibitem[Harding and Powis(2006)Harding, and Powis]{Harding_carvone_CMSXa_2006}
Harding,~C.~J.; Powis,~I. Sensitivity of photoelectron circular dichroism to
  structure and electron dynamics in the photoionization of carvone and related
  chiral monocyclic terpenone enantiomers. \emph{J. Chem. Phys.} \textbf{2006},
  \emph{125}, 234306\relax
\mciteBstWouldAddEndPuncttrue
\mciteSetBstMidEndSepPunct{\mcitedefaultmidpunct}
{\mcitedefaultendpunct}{\mcitedefaultseppunct}\relax
\EndOfBibitem
\bibitem[Rouquet \latin{et~al.}(2023)Rouquet, Roy~Chowdhury, Garcia, Nahon,
  Dupont, Lepère, Le~Barbu-Debus, and Zehnacker]{rouquet_inducedPECD_2023}
Rouquet,~E.; Roy~Chowdhury,~M.; Garcia,~G.~A.; Nahon,~L.; Dupont,~J.;
  Lepère,~V.; Le~Barbu-Debus,~K.; Zehnacker,~A. Induced photoelectron circular
  dichroism onto an achiral chromophore. \emph{Nat. Commun.} \textbf{2023},
  \emph{14}, 6290\relax
\mciteBstWouldAddEndPuncttrue
\mciteSetBstMidEndSepPunct{\mcitedefaultmidpunct}
{\mcitedefaultendpunct}{\mcitedefaultseppunct}\relax
\EndOfBibitem
\bibitem[Losada and Xu(2007)Losada, and Xu]{losada_vcd-h2o_2007}
Losada,~M.; Xu,~Y. Chirality transfer through hydrogen-bonding: {Experimental}
  and ab initio analyses of vibrational circular dichroism spectra of methyl
  lactate in water. \emph{Phys. Chem. Chem. Phys.} \textbf{2007}, \emph{9},
  3127--3135\relax
\mciteBstWouldAddEndPuncttrue
\mciteSetBstMidEndSepPunct{\mcitedefaultmidpunct}
{\mcitedefaultendpunct}{\mcitedefaultseppunct}\relax
\EndOfBibitem
\bibitem[Debie \latin{et~al.}(2008)Debie, Jaspers, Bultinck, Herrebout, and Van
  Der~Veken]{debie_induced-VCD_2008}
Debie,~E.; Jaspers,~L.; Bultinck,~P.; Herrebout,~W.; Van Der~Veken,~B. Induced
  solvent chirality: {A} {VCD} study of camphor in {CDCl3}. \emph{Chem. Phys.
  Lett.} \textbf{2008}, \emph{450}, 426--430\relax
\mciteBstWouldAddEndPuncttrue
\mciteSetBstMidEndSepPunct{\mcitedefaultmidpunct}
{\mcitedefaultendpunct}{\mcitedefaultseppunct}\relax
\EndOfBibitem
\bibitem[Yang and Xu(2009)Yang, and Xu]{yang_vcd-h2o-2_2009}
Yang,~G.; Xu,~Y. Probing chiral solute-water hydrogen bonding networks by
  chirality transfer effects: {A} vibrational circular dichroism study of
  glycidol in water. \emph{J. Chem. Phys.} \textbf{2009}, \emph{130},
  164506\relax
\mciteBstWouldAddEndPuncttrue
\mciteSetBstMidEndSepPunct{\mcitedefaultmidpunct}
{\mcitedefaultendpunct}{\mcitedefaultseppunct}\relax
\EndOfBibitem
\bibitem[Jähnigen \latin{et~al.}(2021)Jähnigen, Sebastiani, and
  Vuilleumier]{jahnigen_non-covalent-induced_2021}
Jähnigen,~S.; Sebastiani,~D.; Vuilleumier,~R. The important role of
  non-covalent interactions for the vibrational circular dichroism of lactic
  acid in aqueous solution. \emph{Phys. Chem. Chem. Phys.} \textbf{2021},
  \emph{23}, 17232--17241\relax
\mciteBstWouldAddEndPuncttrue
\mciteSetBstMidEndSepPunct{\mcitedefaultmidpunct}
{\mcitedefaultendpunct}{\mcitedefaultseppunct}\relax
\EndOfBibitem
\bibitem[Konstantinovsky \latin{et~al.}(2022)Konstantinovsky, Perets, Santiago,
  Velarde, Hammes-Schiffer, and Yan]{konstantinovsky_chiralSFG_2022}
Konstantinovsky,~D.; Perets,~E.~A.; Santiago,~T.; Velarde,~L.;
  Hammes-Schiffer,~S.; Yan,~E. C.~Y. Detecting the {First} {Hydration} {Shell}
  {Structure} around {Biomolecules} at {Interfaces}. \emph{ACS Cent. Sci.}
  \textbf{2022}, \emph{8}, 1404--1414\relax
\mciteBstWouldAddEndPuncttrue
\mciteSetBstMidEndSepPunct{\mcitedefaultmidpunct}
{\mcitedefaultendpunct}{\mcitedefaultseppunct}\relax
\EndOfBibitem
\bibitem[Wang and Cann(2008)Wang, and Cann]{wang_MC_inducedCD_2008}
Wang,~S.; Cann,~N.~M. A molecular dynamics study of chirality transfer: {The}
  impact of a chiral solute on an achiral solvent. \emph{J. Chem. Phys.}
  \textbf{2008}, \emph{129}, 054507\relax
\mciteBstWouldAddEndPuncttrue
\mciteSetBstMidEndSepPunct{\mcitedefaultmidpunct}
{\mcitedefaultendpunct}{\mcitedefaultseppunct}\relax
\EndOfBibitem
\bibitem[Eppink and Parker(1997)Eppink, and Parker]{eppinkparker_vmi_1997}
Eppink,~A.~T.; Parker,~D.~H. Velocity map imaging of ions and electrons using
  electrostatic lenses: {Application} in photoelectron and photofragment ion
  imaging of molecular oxygen. \emph{Rev. Sci. Instrum.} \textbf{1997},
  \emph{68}, 3477--3484\relax
\mciteBstWouldAddEndPuncttrue
\mciteSetBstMidEndSepPunct{\mcitedefaultmidpunct}
{\mcitedefaultendpunct}{\mcitedefaultseppunct}\relax
\EndOfBibitem
\bibitem[Koralek \latin{et~al.}(2018)Koralek, Kim, Bruza, Curry, Chen, Bechtel,
  Cordones, Sperling, Toleikis, Kern, Moeller, Glenzer, and
  DePonte]{Koralek_ultrathin-FJ_2018}
Koralek,~J.~D.; Kim,~J.~B.; Bruza,~P.; Curry,~C.~B.; Chen,~Z.; Bechtel,~H.~A.;
  Cordones,~A.~A.; Sperling,~P.; Toleikis,~S.; Kern,~J.~F.; Moeller,~S.~P.;
  Glenzer,~S.~H.; DePonte,~D.~P. Generation and characterization of ultrathin
  free-flowing liquid sheets. \emph{Nat. Commun.} \textbf{2018}, \emph{9},
  1353\relax
\mciteBstWouldAddEndPuncttrue
\mciteSetBstMidEndSepPunct{\mcitedefaultmidpunct}
{\mcitedefaultendpunct}{\mcitedefaultseppunct}\relax
\EndOfBibitem
\bibitem[Viefhaus \latin{et~al.}(2013)Viefhaus, Scholz, Deinert, Glaser,
  Ilchen, Seltmann, Walter, and Siewert]{viefhaus_P04_2013}
Viefhaus,~J.; Scholz,~F.; Deinert,~S.; Glaser,~L.; Ilchen,~M.; Seltmann,~J.;
  Walter,~P.; Siewert,~F. The {Variable} {Polarization} {XUV} {Beamline} {P04}
  at {PETRA} {III}: {Optics}, mechanics and their performance. \emph{Nucl.
  Instrum. Methods Phys. Res. A} \textbf{2013}, \emph{710}, 151--154\relax
\mciteBstWouldAddEndPuncttrue
\mciteSetBstMidEndSepPunct{\mcitedefaultmidpunct}
{\mcitedefaultendpunct}{\mcitedefaultseppunct}\relax
\EndOfBibitem
\end{mcitethebibliography}
\makeatletter\@input{xx2.tex}\makeatother
\end{document}

\singlespacing

\section{Background Subtraction and Data Analysis}
We attempted many approaches to background subtraction, including the use of an experimental background measured for neat water solutions without alanine under the same experimental conditions for which we measured the corresponding alanine spectra. However, we found that such experimentally measured background spectra provided an imperfect fit to the data, and we were unable to sufficiently improve the fit through simple scaling of the spectra. In general, the measurement of an appropriate background spectrum for low-KE PES of condensed-phase targets is not straightforward. Although the simple measurement of a reference sample (e.g. pure water) would seem to provide a reasonable background, the measured low-energy electron background spectrum of any sample will naturally strongly depend on the integral ionization cross section of all species contained in the sample, their surface propensities, and the specific electron mean free path within that sample. These effects combined to make the measurement of an appropriate experimental background for kinetic energies below 15--20~eV very challenging. We also attempted to measure experimental background spectra by keeping the solution unchanged, and instead increasing the photon energy sufficiently such that the alanine C~1s features were pushed to higher kinetic energy and out of the measurement range. This approach also proved unsuitable, as all of the previously listed factors depend critically on photon energy. Ultimately, we chose to fit the background using a variety of functions and determined that polynomials and sums of polynomials and exponentials constituted the most effective and reproducible background function for these spectra.

\begin{figure}
	\includegraphics[width=\textwidth]{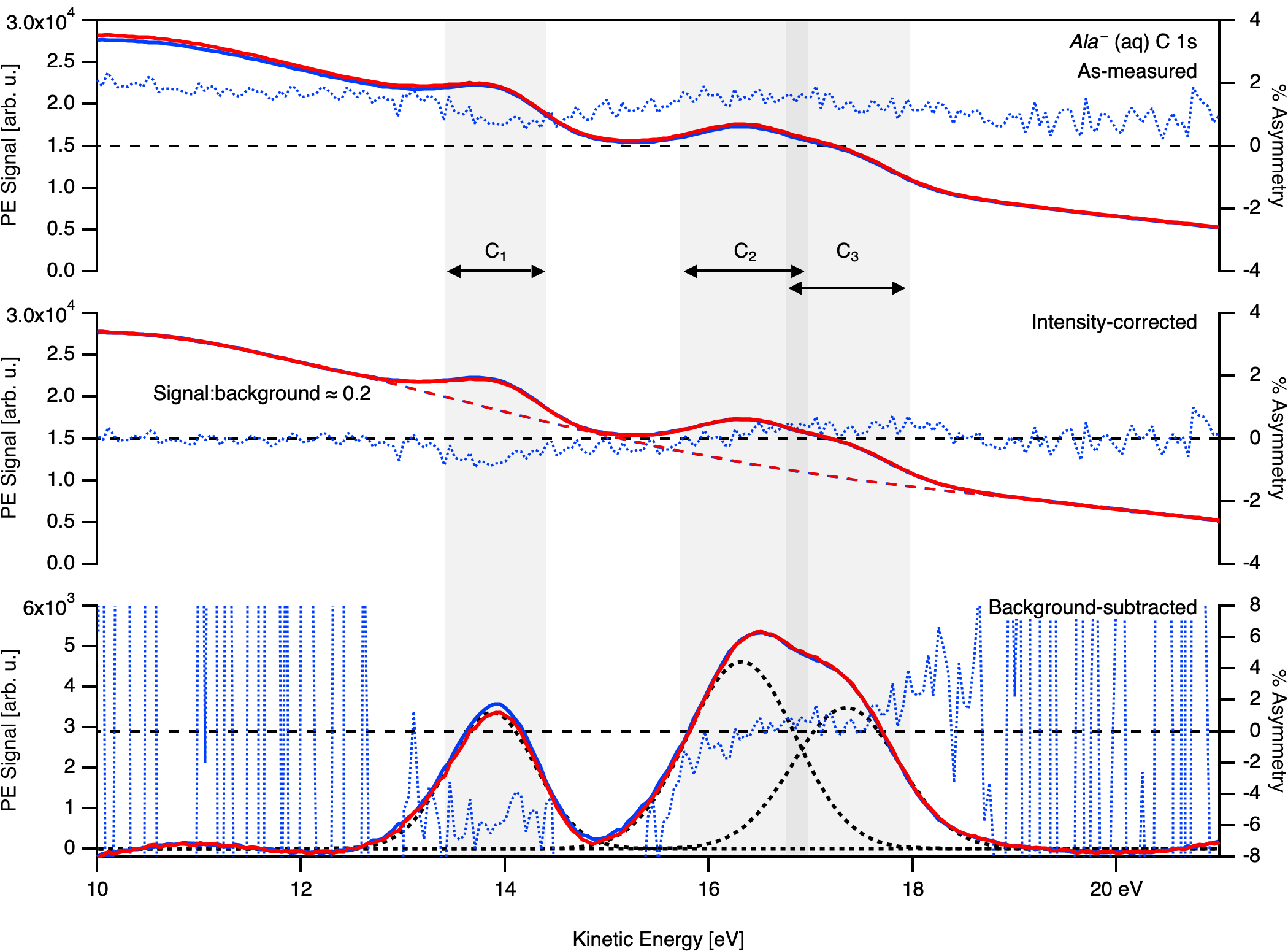}
	\caption{Example of data analysis workflow for a sample pair of spectra measured for $Ala^{-}$ with h$\nu$=307~eV photons. Top: As-measured spectra measured with left- and right-handed circularly polarized light (blue and red solid lines, respectively), with calculated asymmetry (dotted blue line). Middle: Intensity-corrected spectra and associated background (dashed red and blue lines), with calculated asymmetry. Bottom: Background-subtracted spectra, with associated C${1}$, C${2}$, and C${3}$ peak fits and calculated asymmetry. The gray areas denote the kinetic-energy range within which the asymmetry values are averaged to produce a single value of $b^{+1}_{1}$. Note that the magnitude of the asymmetry calculated after background subtraction is significantly larger than that calculated prior to background subtraction, reflecting the low signal-to-background ratio of the as-measured data.}
	\label{fig:S0}
\end{figure}

\begin{figure}
	\center
	\includegraphics[width=\textwidth]{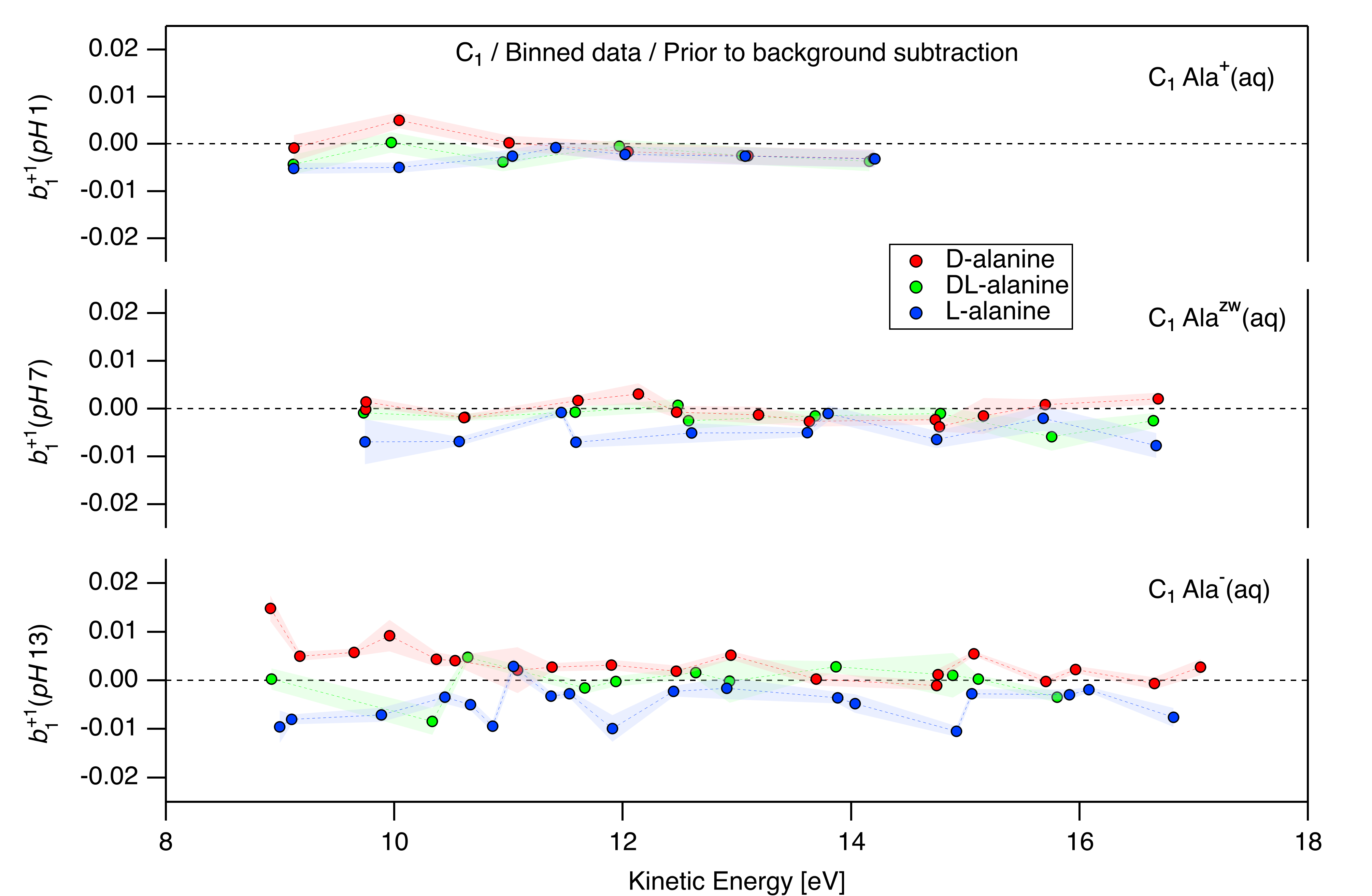}
	\caption{Values of the $b_{1}^{+1}$ photoionization parameter obtained for C~1s measurements of aqueous solutions of D-, L-, and DL-alanine (red, blue, and green points, respectively) at pH 1, 7, and 13 (top, middle, and bottom; corresponding to the cationic, zwitterionic, and anionic form of the molecule, respectively). All $b_{1}^{+1}$ values shown correspond to photoionization of the C$_{1}$ carboxylic acid group prior to background subtraction. The data is displayed with a kinetic-energy binning of 250~meV.}  
	\label{fig:S4}
\end{figure}

\begin{figure}
	\center
	\includegraphics[width=\textwidth]{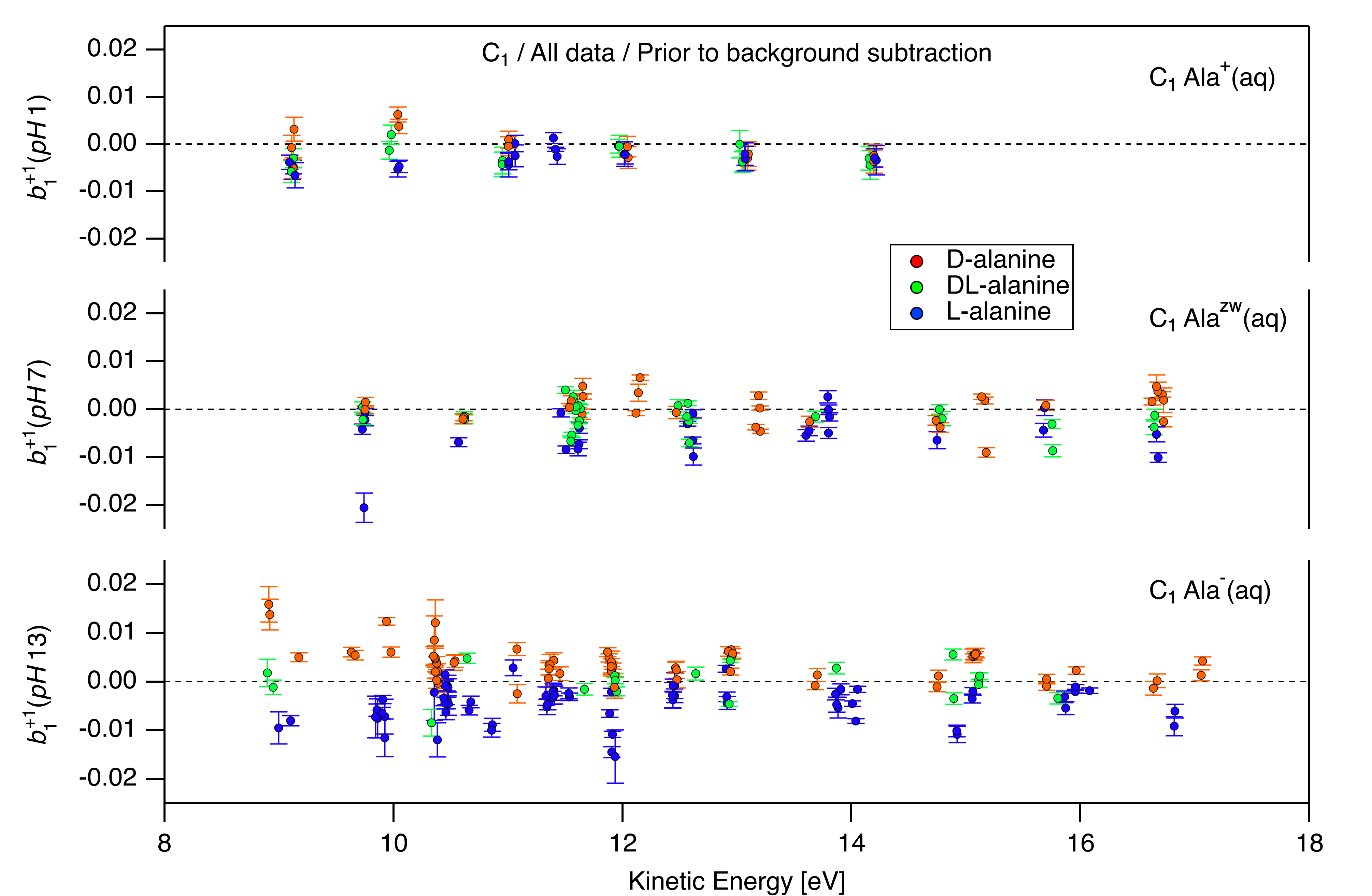}
	\caption{Values of the $b_{1}^{+1}$ photoionization parameter obtained for C~1s measurements of aqueous solutions of D-, L-, and DL-alanine (red, blue, and green points, respectively) at pH 1, 7, and 13 (top, middle, and bottom; corresponding to the cationic, zwitterionic, and anionic form of the molecule, respectively). All $b_{1}^{+1}$ values shown correspond to photoionization of the C$_{1}$ carboxylic acid group prior to background subtraction.}  
	\label{fig:S5}
\end{figure}

\begin{figure}
	\center
	\includegraphics[width=\textwidth]{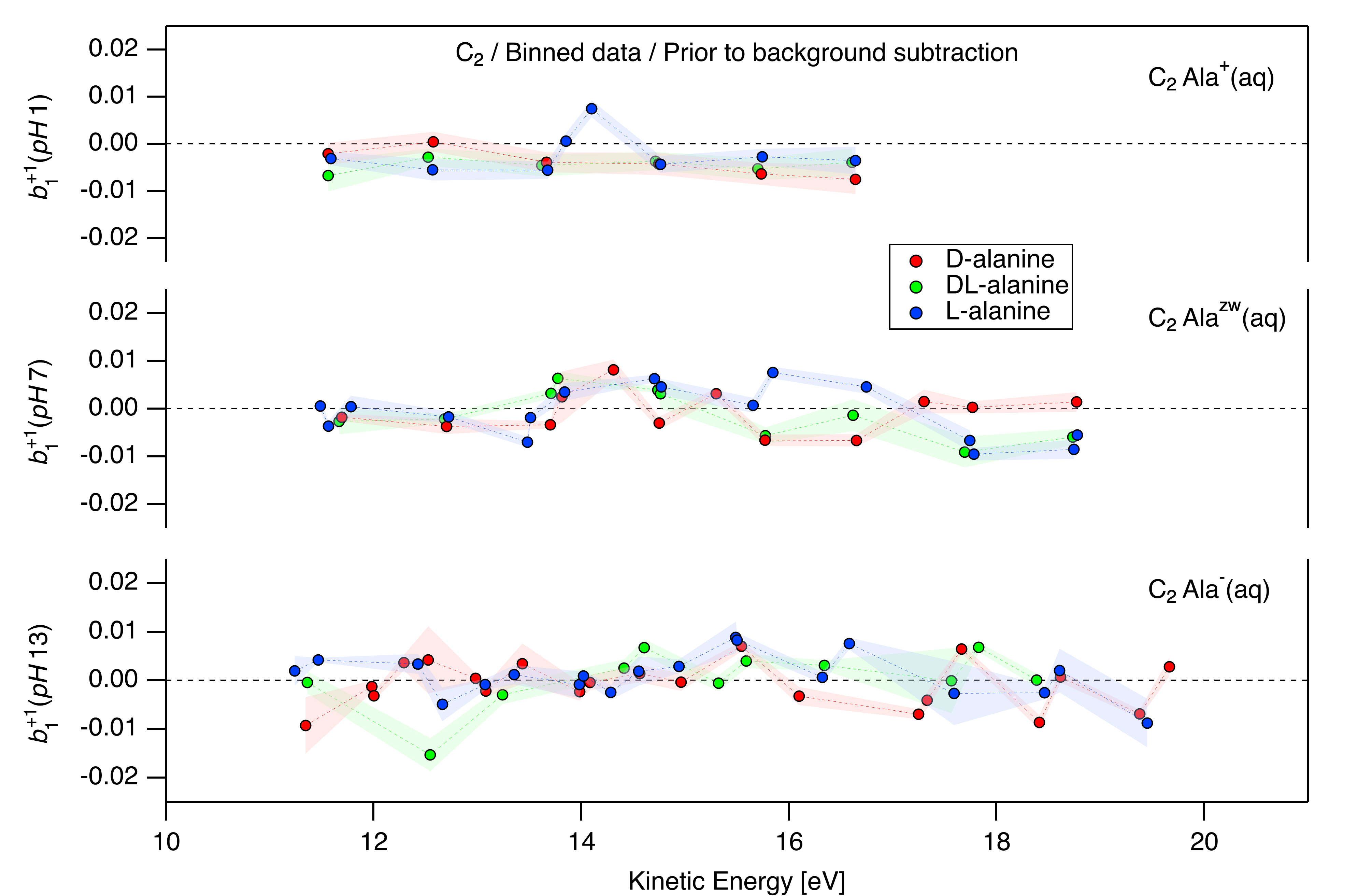}
	\caption{Values of the $b_{1}^{+1}$ photoionization parameter obtained for C~1s measurements of aqueous solutions of D-, L-, and DL-alanine (red, blue, and green points, respectively) at pH 1, 7, and 13 (top, middle, and bottom; corresponding to the cationic, zwitterionic, and anionic form of the molecule, respectively). All $b_{1}^{+1}$ values shown correspond to photoionization of the C$_{2}$ chiral center prior to background subtraction. The data is displayed with a kinetic-energy binning of 250~meV.}  
	\label{fig:S8}
\end{figure}

\begin{figure}
	\center
	\includegraphics[width=\textwidth]{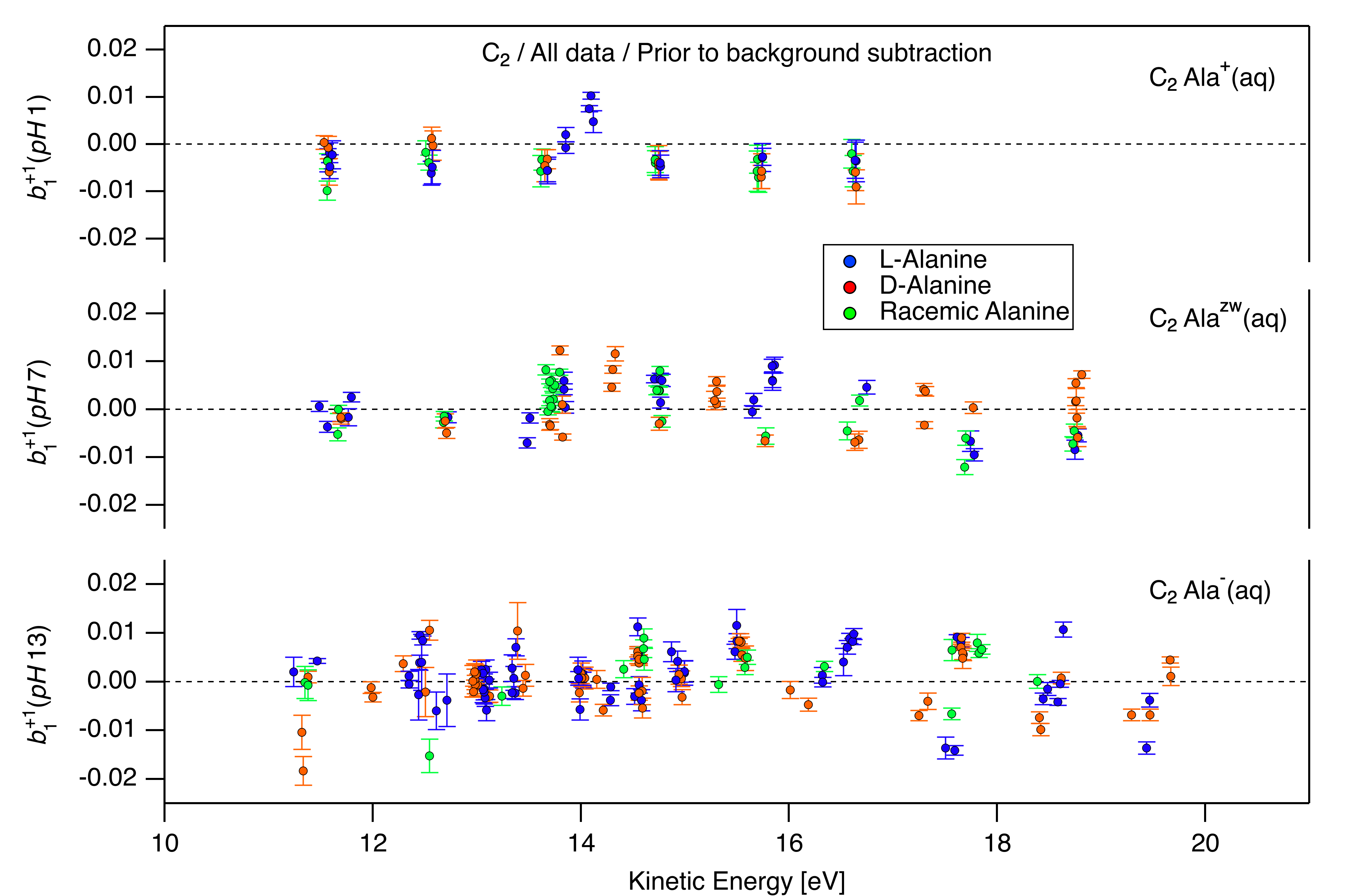}
	\caption{Values of the $b_{1}^{+1}$ photoionization parameter obtained for C~1s measurements of aqueous solutions of D-, L-, and DL-alanine (red, blue, and green points, respectively) at pH 1, 7, and 13 (top, middle, and bottom; corresponding to the cationic, zwitterionic, and anionic form of the molecule, respectively). All $b_{1}^{+1}$ values shown correspond to photoionization of the C$_{2}$ chiral center prior to background subtraction.}  
	\label{fig:S9}
\end{figure}

\begin{figure}
	\center
	\includegraphics[width=\textwidth]{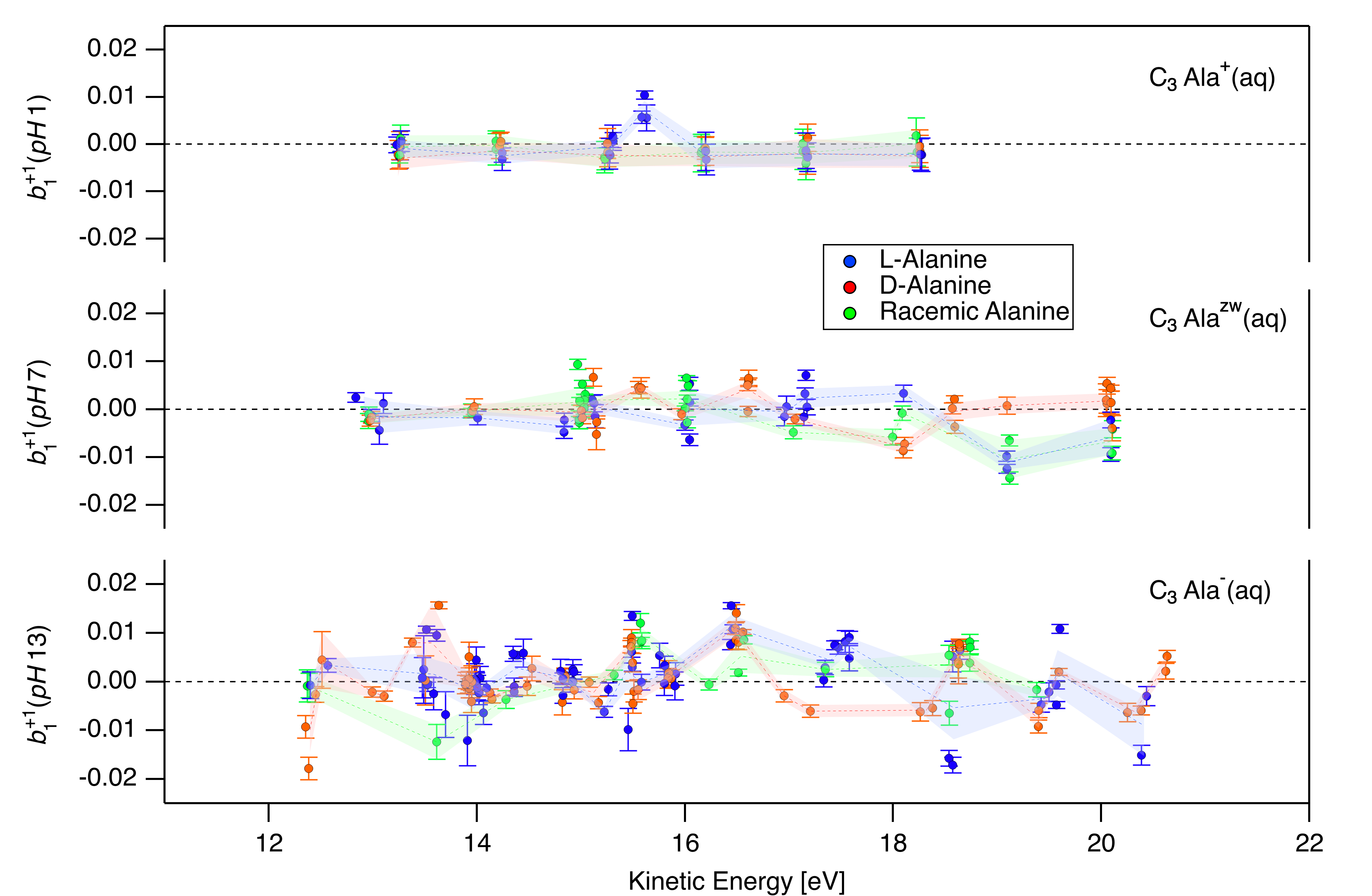}
	\caption{Values of the $b_{1}^{+1}$ photoionization parameter obtained for C~1s measurements of aqueous solutions of D-, L-, and DL-alanine (red, blue, and green points, respectively) at pH 1, 7, and 13 (top, middle, and bottom; corresponding to the cationic, zwitterionic, and anionic form of the molecule, respectively). All $b_{1}^{+1}$ values shown correspond to photoionization of the C$_{3}$ methyl group prior to background subtraction. The data is displayed with a kinetic-energy binning of 250~meV.}  
	\label{fig:S12}
\end{figure}

\begin{figure}
	\center
	\includegraphics[width=\textwidth]{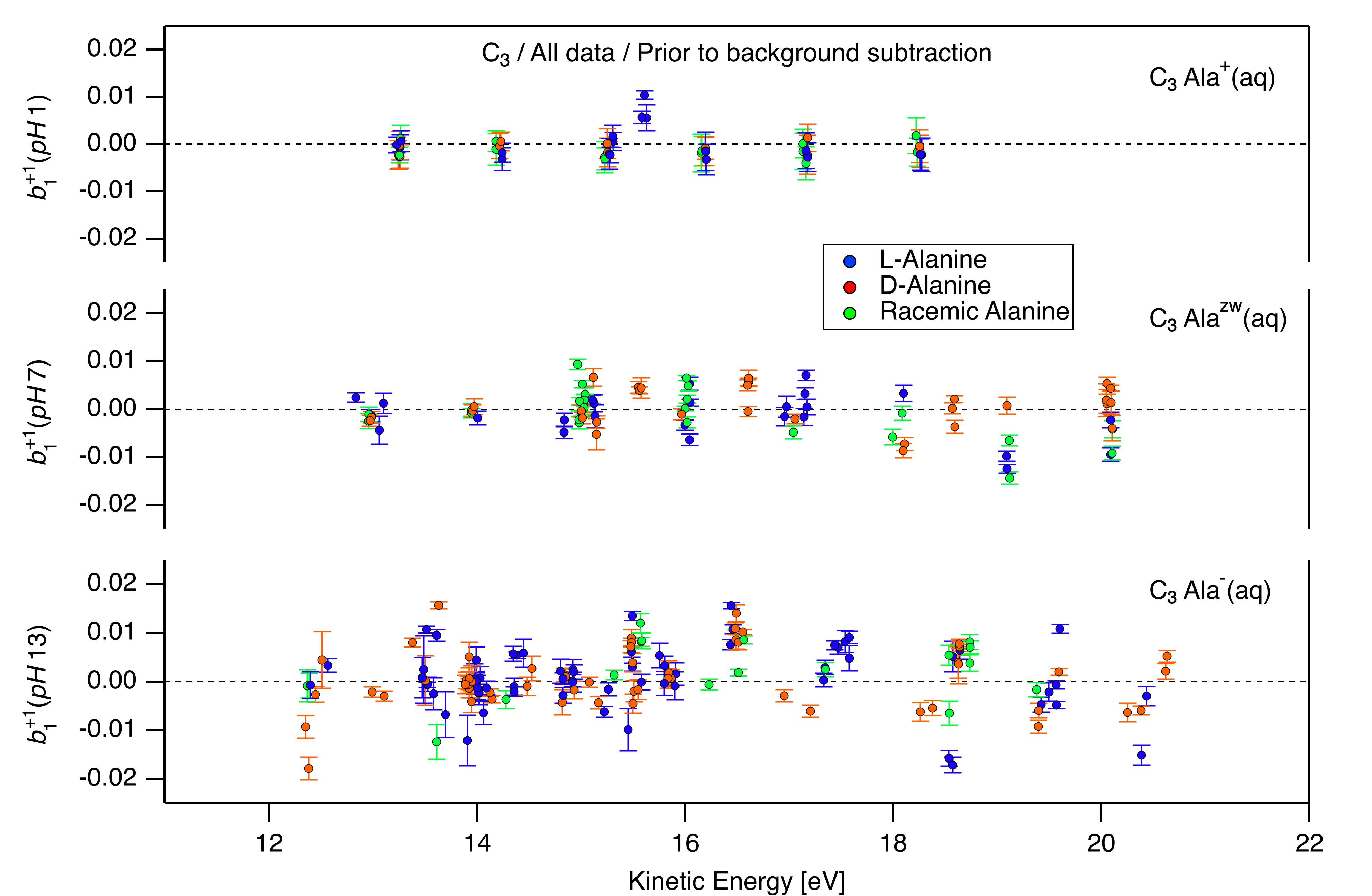}
	\caption{Values of the $b_{1}^{+1}$ photoionization parameter obtained for C~1s measurements of aqueous solutions of D-, L-, and DL-alanine (red, blue, and green points, respectively) at pH 1, 7, and 13 (top, middle, and bottom; corresponding to the cationic, zwitterionic, and anionic form of the molecule, respectively). All $b_{1}^{+1}$ values shown correspond to photoionization of the C$_{3}$ methyl group prior to background subtraction. }  
	\label{fig:S13}
\end{figure}

Data were analyzed pairwise and for each pair of intensity-corrected spectra, we manually defined a region of interest containing all three of the C 1s peaks and performed a fit to the background with this region masked. We were careful to avoid overfitting, and only accepted a fitted background if small changes to the masked region did not greatly influence the resulting peak shapes and intensities. We then calculated the point-wise difference-over-sum for the $+$ and $-$ data as appears in Eq.~\ref{Maintext-Equation 3}. This was repeated for the data prior to background subtraction; these data served as a guide during background subtraction and are presented below (Figs.~\ref{fig:S4}, \ref{fig:S5}, \ref{fig:S8}, \ref{fig:S9}, \ref{fig:S12}, and \ref{fig:S13}). We identified the peak positions in the background-subtracted data by locating maxima in the second derivatives of the spectra. For each peak, we defined a kinetic-energy window centered at the peak position with a width defined by the full width at half maximum of the peak. Minor differences in peak positions due to offsets in photon energy between left- and right-handed CPL beamline settings were usually less than 5--10~meV and were corrected manually prior to determination of $b^{+1}_{1}$. We then calculated the mean and standard deviation for the values of the asymmetry arising from the data within these energy windows. These asymmetries were subsequently multiplied by a geometric factor taking into account the measurement angle of 50$^{\circ}$, resulting in a single value of $b^{+1}_{1}$ per peak based on Eq.~\ref{Maintext-Equation 2}. This process is illustrated in Fig.~\ref{fig:S0} 

A single datapoint, as presented in Fig.~\ref{fig:S4}, thus constitutes the comparison between a pair of spectra measured with left- and right-handed CPL. Each spectrum used for such analysis is in general taken as an average of 10 acquisitions. The time investment for a single data point is approximately one hour of measurement time.

%Ke-shift.

\section{Electronic Circular Dichroism}
\begin{figure}
	\includegraphics[width=0.5\textwidth]{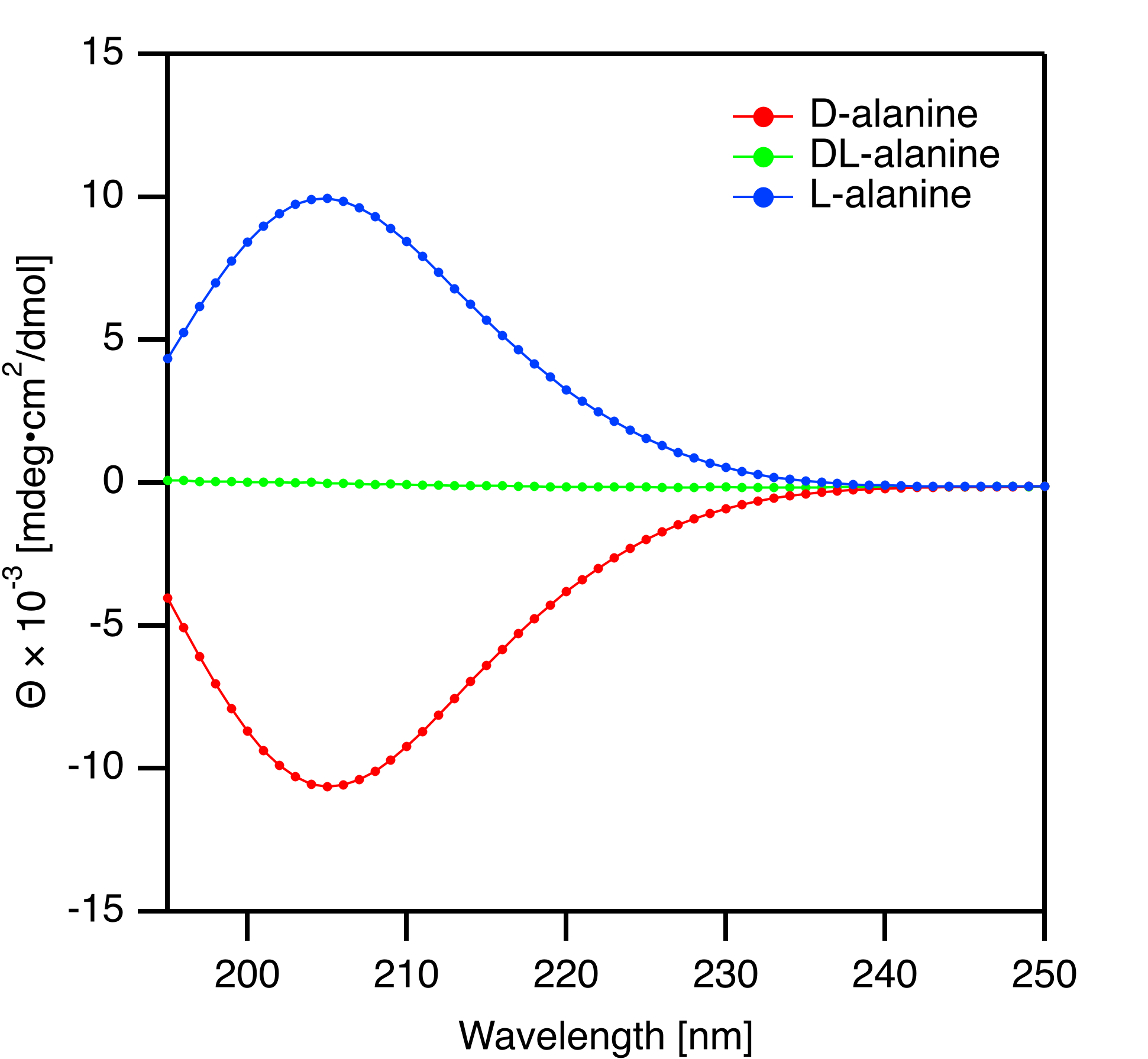}
	\caption{Electronic circular dichroism absorption spectra for 50~mM aqueous solutions of  D-, L-, and DL-alanine (red, blue, and green lines, respectively). The data demonstrate clear optical activity only for the enantiopure samples. All data are plotted in units of molar ellipticity.}
	\label{fig:S1}
\end{figure}

Absorption-based electronic circular-dichroism measurements were performed using a commercial Chirascan 100 CD spectrophotometer (Applied Photophysics). Dilute (50 mM) aqueous solutions of D-, L-, and DL-alanine were prepared using ultrapure water without pH adjustment, and measured in a quartz cuvette with a 1~mm path length. Electronic circular dichroism spectra were recorded in the wavelength range of 195--250~nm with a step size of 1~nm. For each sample, we measured 3 spectra. The averaged results are presented in Fig.~\ref{fig:S1}

\section{Calculation of Alanine $b^{0}_{2}$ Photoionization Parameter}

Calculations to predict the $b^{0}_{2}$ angular distribution parameter for C~1s photoemission from alanine were made using the CMS-X$\alpha$ method as fully described and applied previously by Tia et al ~\cite{tia_alaPECD2_2014}. Those calculations examined the isolated (gas-phase) neutral molecule’s valence photoionization but were here extended to treat core-level photoemission from the three carbon sites. Three different conformers of alanine were modeled, believed to be the most stable configurations of the gas-phase molecule. Conformers 1, 2, and 3 correspond to those descriptions used in reference~\citenum{tia_alaPECD2_2014}. In the kinetic-energy range investigated, the $\beta$ parameter for core-level photoionization was found to display little conformer dependence.

\begin{figure}
	\center
	\includegraphics[width=\textwidth]{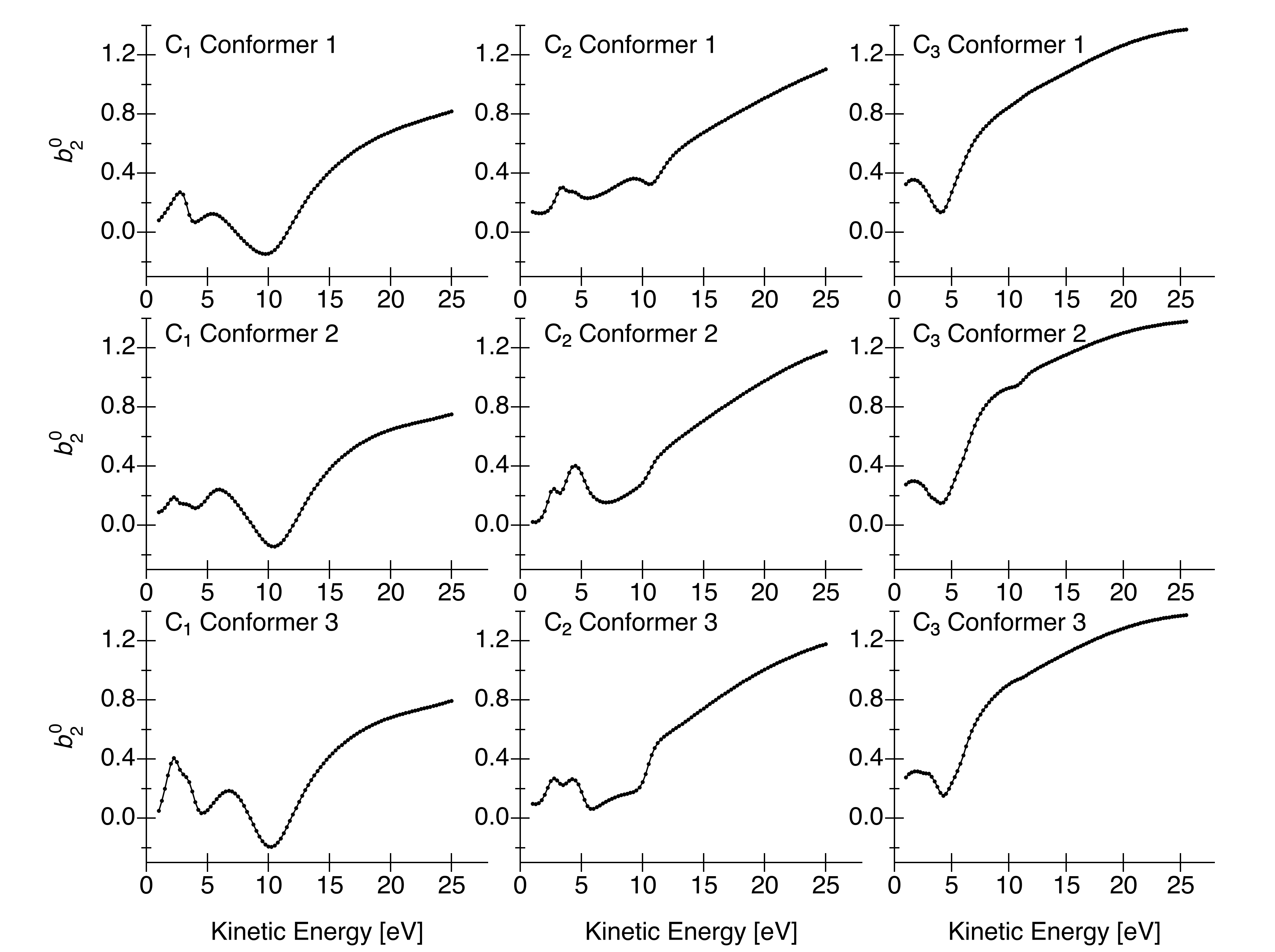}
	\caption{Calculated values of $b^{0}_{2}$ for 1s photoionization of alanine's three carbon groups as a function of photoelectron kinetic energy. C$_{1}$, C$_{2}$, and C$_{3}$ correspond to alanine's carboxylic group, chiral center, and methyl group, respectively, as described in the main text. Conformers 1, 2, and 3 are the same as those discussed in reference \citenum{tia_alaPECD2_2014}.}
	\label{fig:S2}
\end{figure}

\section{Photoelectron Circular Dichroism Data for all Carbon Groups}

Although the primary focus of our study was on photoelectron circular dichroism in photoionization of alanine's COOH/COO$^{-}$ group, we also collected data for alanine's other carbon centers. Here, we include data corresponding to photoionization of alanine's chiral center (C$_{2}$) and methyl group (C$_{3}$), as well as values for $b_{1}^{+}$ based only on the measured data prior to background subtraction.

\subsubsection*{Error Propagation for Binned Datasets}
For clarity, we have displayed the data in the main text following binning of data points using a 250 meV kinetic-energy window. The shaded regions in these plots represent the combined error of the individual data points as well as the spread of the data points binned. The error of each data point was calculated as the standard deviation of the percent difference across the full width at half maximum of the PE feature (see Fig. \ref{Maintext-fig:BGSub}). When $n$ data points were binned together, the error was calculated as: 

\begin{equation}
	Error = \sqrt{\frac{\sum_{i=1}^{n}\sigma^{2}_{i}}{n^{2}}  + \left(\frac{\sigma_{bin}}{\sqrt{n}}\right)^2} ,
\end{equation}

\noindent with $\sigma_{i}$ being the error of a given point, and $\sigma_{bin}$ the standard deviation of the $n$ values of $b^{+1}_{1}$ in a given bin. The relation propagates the individual error of each data point and the standard error of the mean of all data points within a binning window.

\begin{figure}
	\center
	\includegraphics[width=\textwidth]{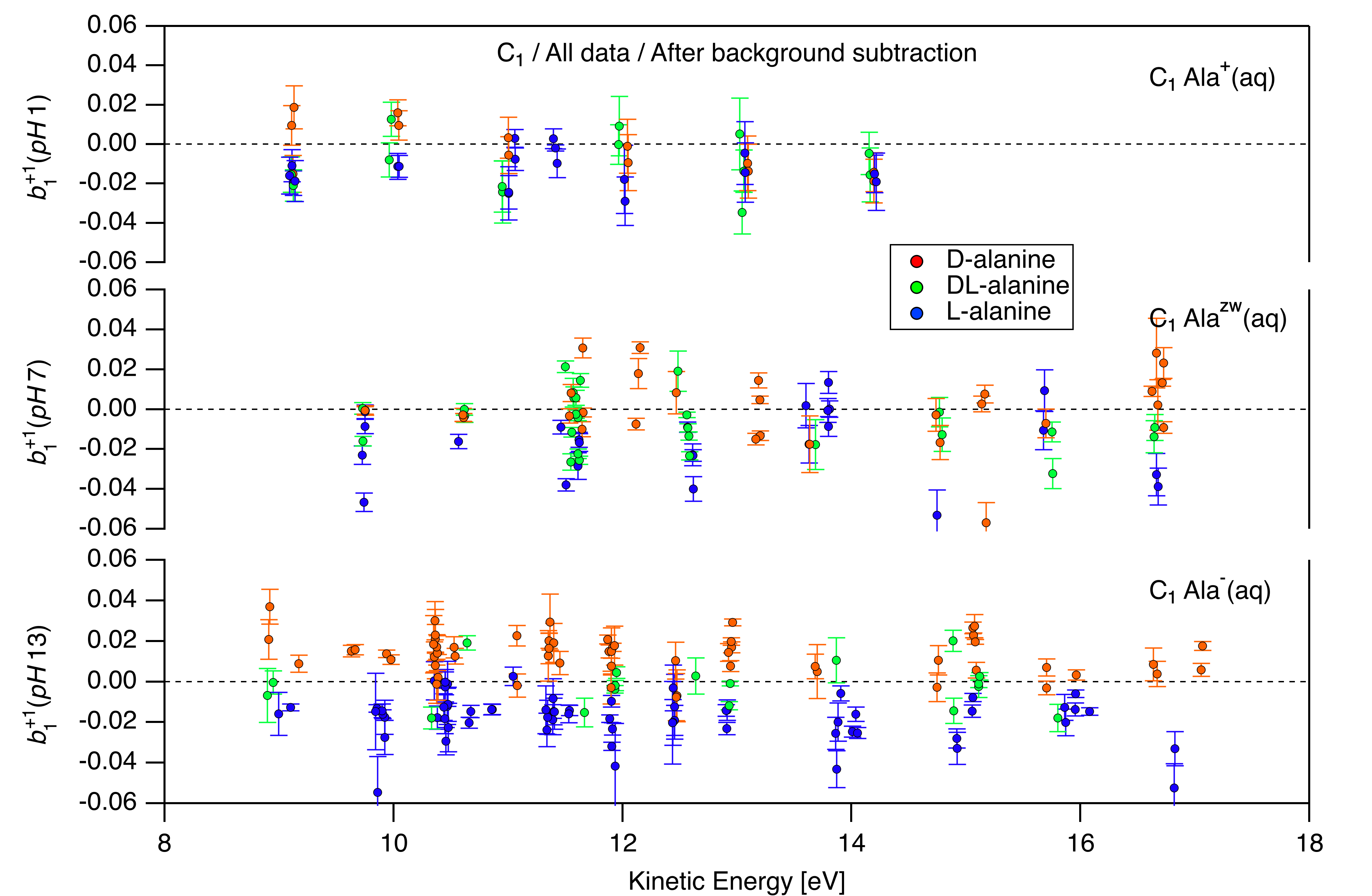}
	\caption{Values of the $b_{1}^{+1}$ photoionization parameter obtained for C~1s measurements of aqueous solutions of D-, L-, and DL-alanine (red, blue, and green points, respectively) at pH 1, 7, and 13 (top, middle, and bottom; corresponding to the cationic, zwitterionic, and anionic form of the molecule, respectively).  All $b_{1}^{+1}$ values shown correspond to photoionization of the C$_{1}$ carboxylic acid group.}  
	\label{fig:S3}
\end{figure}

\begin{figure}
	\center
	\includegraphics[width=\textwidth]{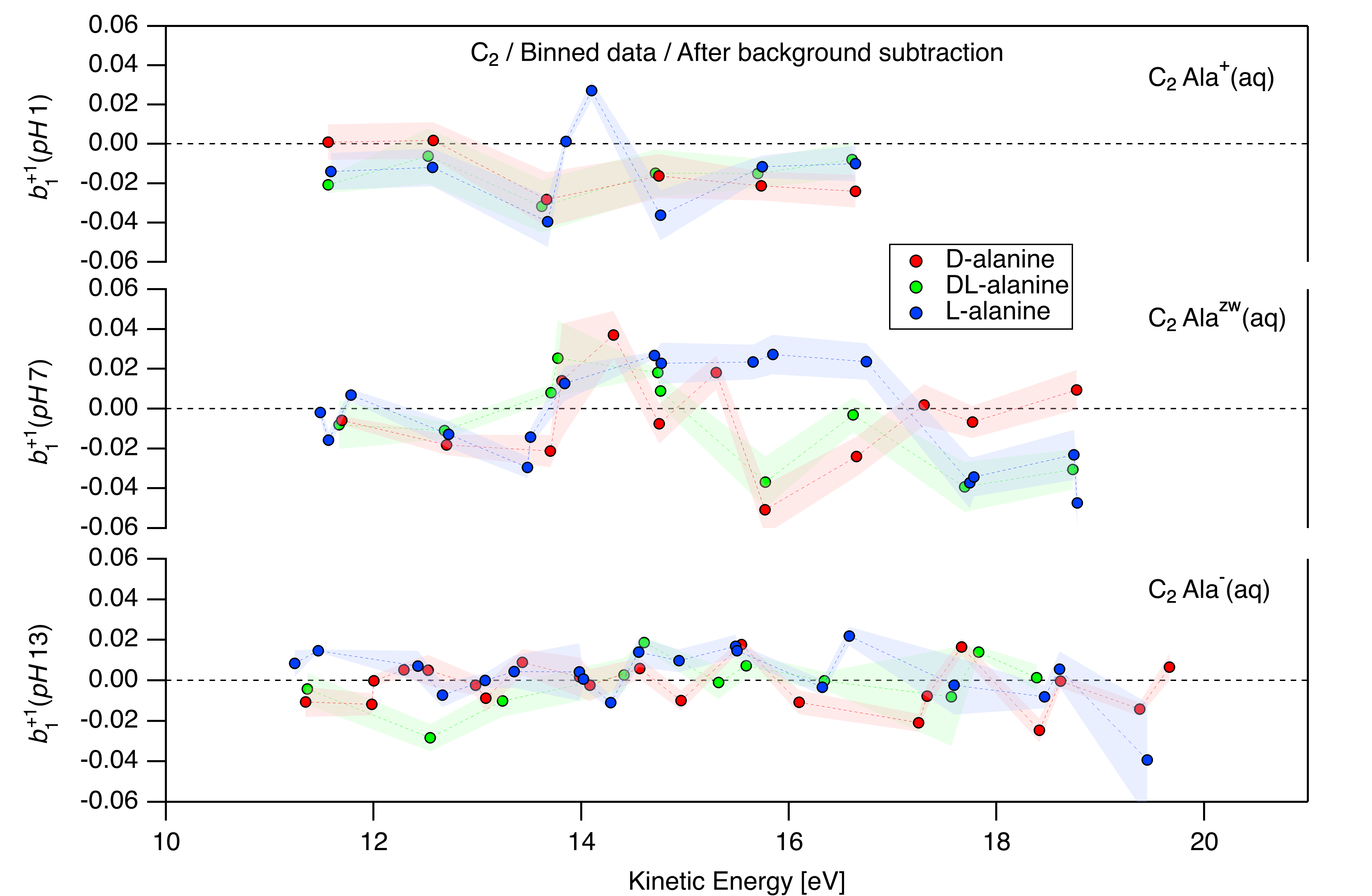}
	\caption{Values of the $b_{1}^{+1}$ photoionization parameter obtained for C~1s measurements of aqueous solutions of D-, L-, and DL-alanine (red, blue, and green points, respectively) at pH 1, 7, and 13 (top, middle, and bottom; corresponding to the cationic, zwitterionic, and anionic form of the molecule, respectively). All $b_{1}^{+1}$ values shown correspond to photoionization of the C$_{2}$ chiral center. The data is displayed with a kinetic-energy binning of 250~meV.}  
	\label{fig:S6}
\end{figure}

\begin{figure}
	\center
	\includegraphics[width=\textwidth]{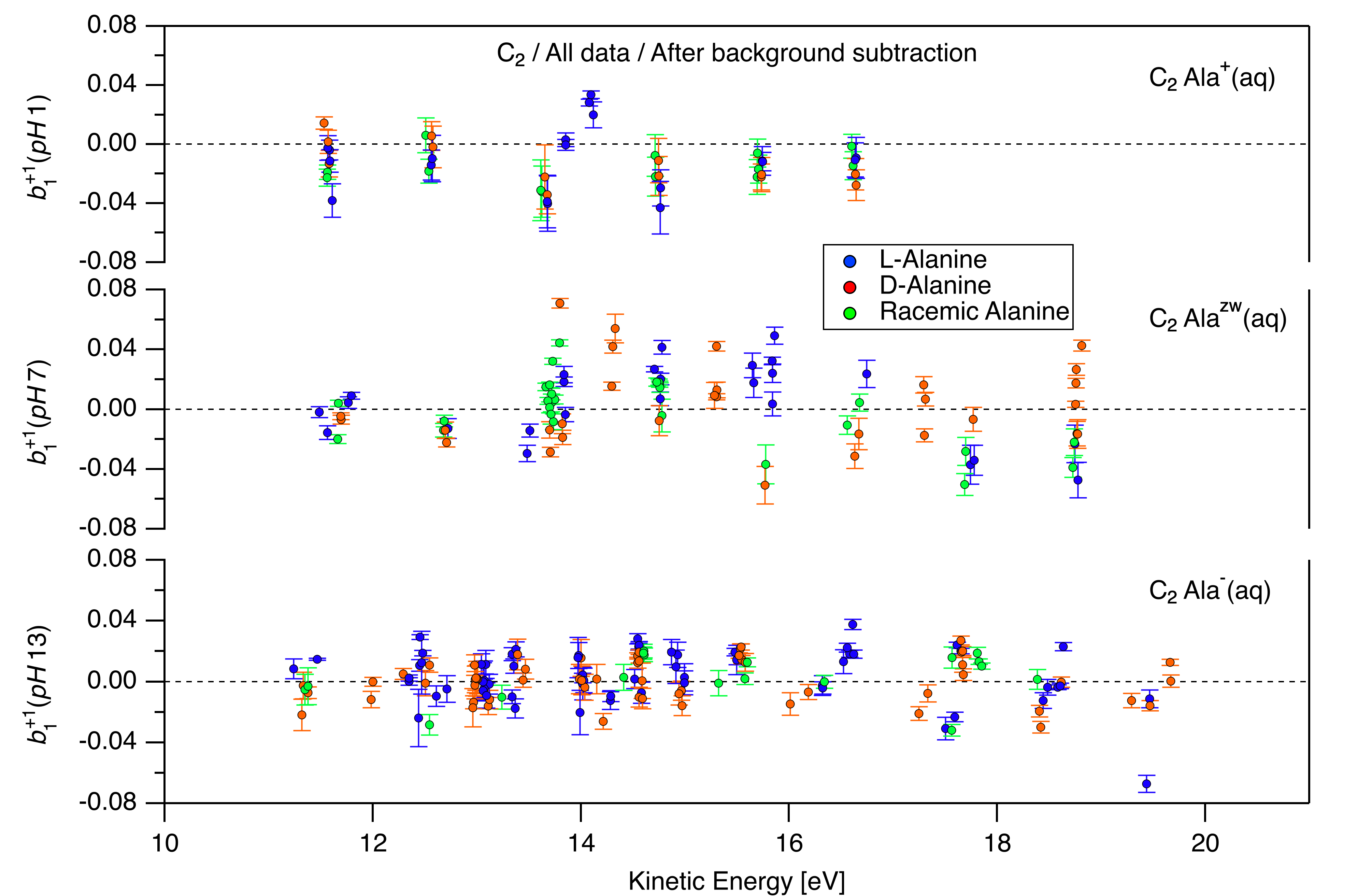}
	\caption{Values of the $b_{1}^{+1}$ photoionization parameter obtained for C~1s measurements of aqueous solutions of D-, L-, and DL-alanine (red, blue, and green points, respectively) at pH 1, 7, and 13 (top, middle, and bottom; corresponding to the cationic, zwitterionic, and anionic form of the molecule, respectively). All $b_{1}^{+1}$ values shown correspond to photoionization of the C$_{2}$ chiral center.}  
	\label{fig:S7}
\end{figure}

\begin{figure}
	\center
	\includegraphics[width=\textwidth]{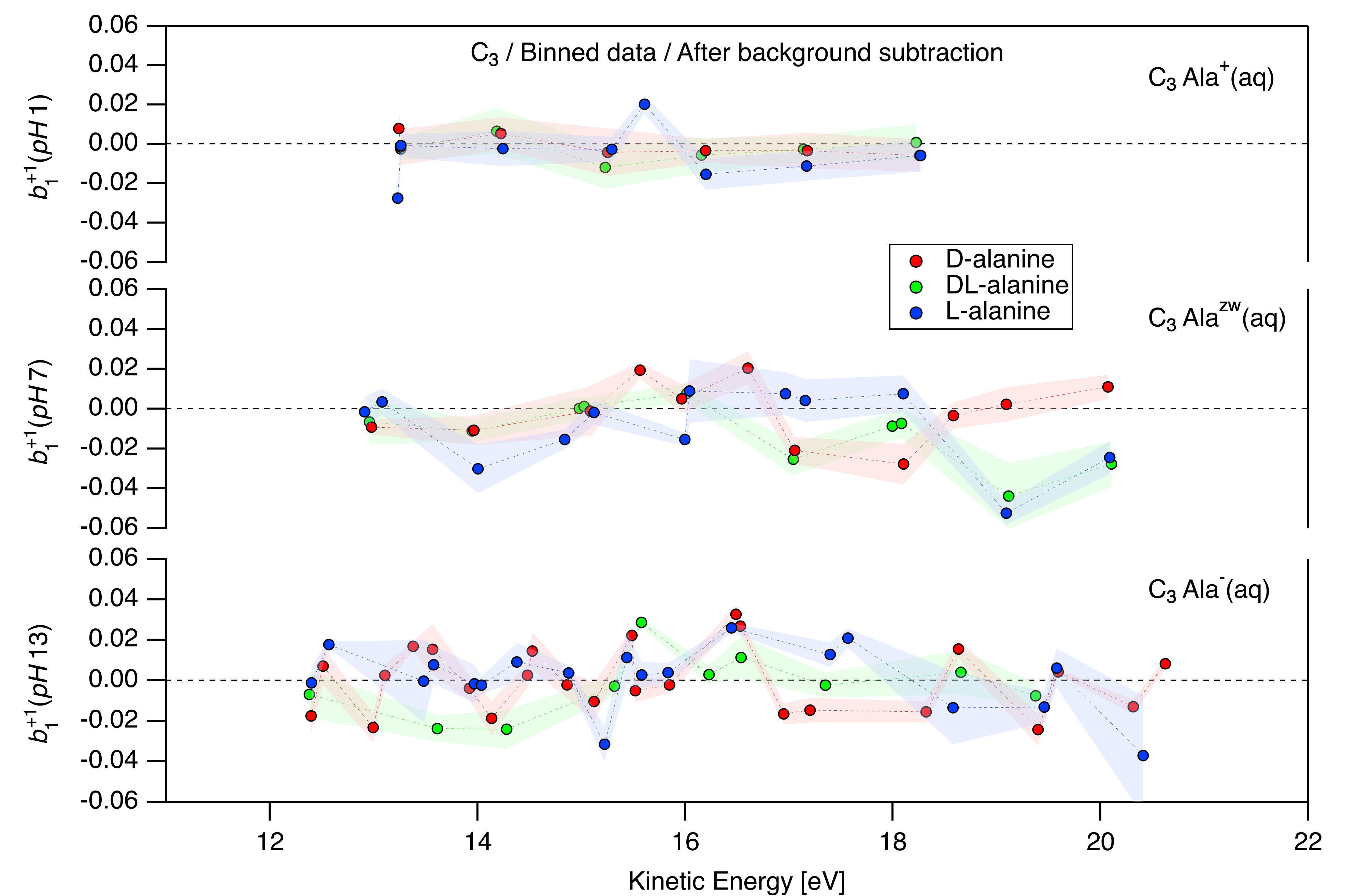}
	\caption{Values of the $b_{1}^{+1}$ photoionization parameter obtained for C~1s measurements of aqueous solutions of D-, L-, and DL-alanine (red, blue, and green points, respectively) at pH 1, 7, and 13 (top, middle, and bottom; corresponding to the cationic, zwitterionic, and anionic form of the molecule, respectively). All $b_{1}^{+1}$ values shown correspond to photoionization of the C$_{3}$ methyl group. The data is displayed with a kinetic-energy binning of 250~meV.}  
	\label{fig:S10}
\end{figure}

\begin{figure}
	\center
	\includegraphics[width=\textwidth]{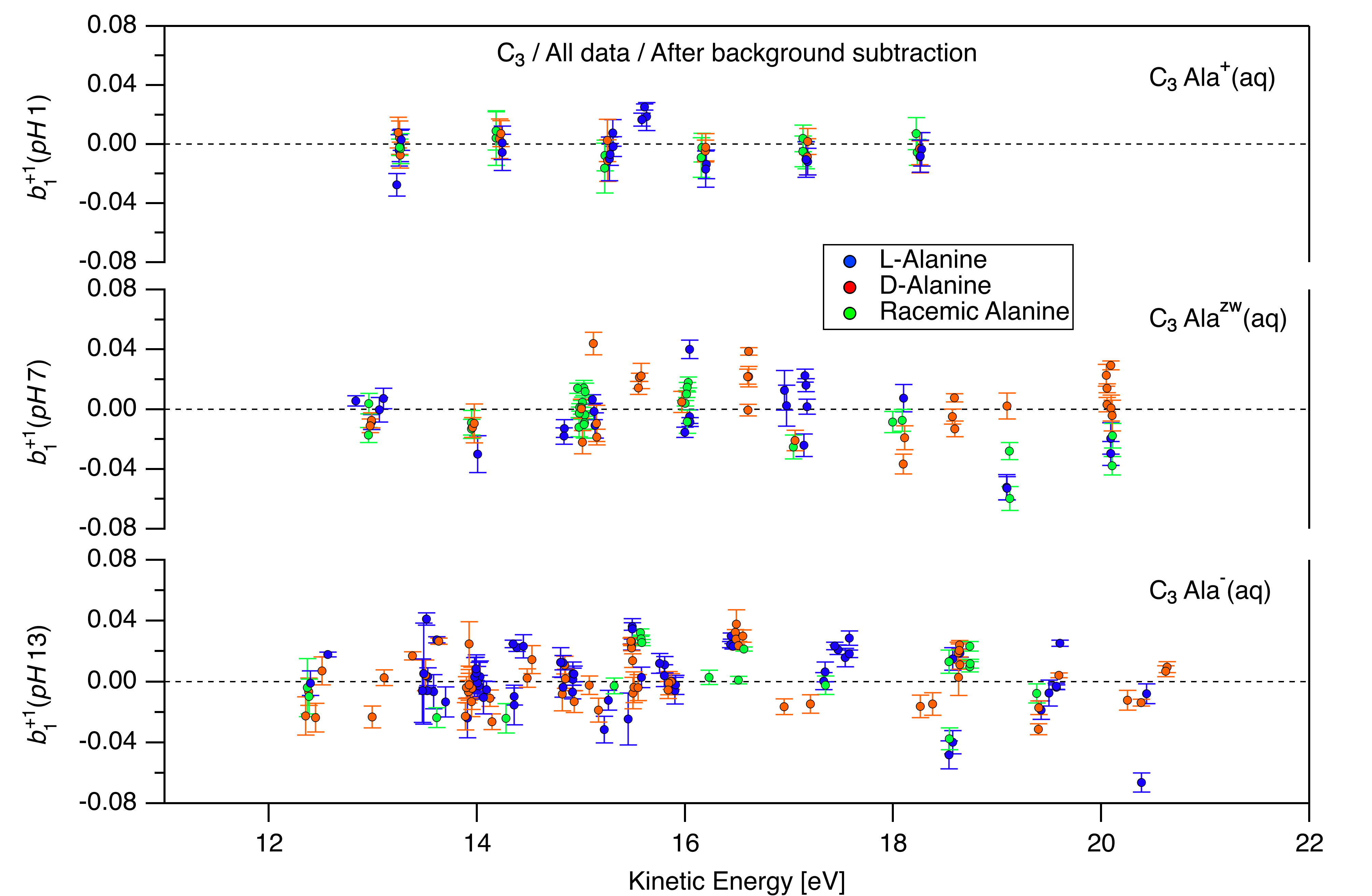}
	\caption{Values of the $b_{1}^{+1}$ photoionization parameter obtained for C~1s measurements of aqueous solutions of D-, L-, and DL-alanine (red, blue, and green points, respectively) at pH 1, 7, and 13 (top, middle, and bottom; corresponding to the cationic, zwitterionic, and anionic form of the molecule, respectively). All $b_{1}^{+1}$ values shown correspond to photoionization of the C$_{3}$ methyl group.}  
	\label{fig:S11}
\end{figure}

\bibliography{PECD_refs.bib}
\makeatletter\@input{xx.tex}\makeatother